\documentclass{article}

\PassOptionsToPackage{numbers,compress}{natbib}
\usepackage[preprint]{neurips_2026}

\usepackage[utf8]{inputenc}
\usepackage[T1]{fontenc}
\usepackage{hyperref}
\usepackage{url}
\usepackage{booktabs}
\usepackage{amsmath}
\usepackage{amssymb}
\usepackage{amsfonts}
\usepackage{microtype}
\usepackage{xcolor}
\usepackage{array}
\usepackage{tabularx}
\usepackage{graphicx}
\graphicspath{{../}}
\usepackage{tikz}
\usepackage{float}
\usepackage{mathrsfs}

\usepackage[dvipsnames,usenames]{xcolor}
\definecolor{linkcolor}{rgb}{0.0,0.3,0.5}

\newcommand{\case}[1]{\texttt{#1}}
\newcommand{\outcome}[1]{\texttt{#1}}
\newcommand{\pp}[1]{\left[\,#1\,\right]_{+}}
\newcommand{\fitdisplay}[1]{\resizebox{\linewidth}{!}{$\displaystyle #1$}}
\newcolumntype{Y}{>{\raggedright\arraybackslash}X}
\definecolor{globalgreen}{HTML}{2E7D32}
\definecolor{nearorange}{HTML}{EF8A17}
\definecolor{localblue}{HTML}{2B6CB0}
\definecolor{rejectpurple}{HTML}{7B3FA1}
\definecolor{badgray}{HTML}{666666}
\definecolor{waterblue}{HTML}{4F8FC9}
\definecolor{wateredge}{HTML}{1A264A}
\definecolor{forbidred}{HTML}{B9413C}
\definecolor{promptbg}{HTML}{F4F8FB}
\definecolor{promptborder}{HTML}{86A9C8}
\definecolor{falsehintbg}{HTML}{FFF4F1}
\definecolor{falsehintborder}{HTML}{D98773}
\definecolor{hintbg}{HTML}{F1F8F2}
\definecolor{hintborder}{HTML}{74A876}
\definecolor{nohintbg}{HTML}{F6F6F6}
\definecolor{nohintborder}{HTML}{8E8E8E}
\definecolor{tracebg}{HTML}{FBFAF2}
\definecolor{traceborder}{HTML}{B9A45B}

\newcommand{\eM}{e_{3}^{-}}
\newcommand{\eP}{e_{3}^{+}}

\newcommand{\promptbox}[2]{%
  \par\smallskip\noindent
  \fcolorbox{promptborder}{promptbg}{%
    \begin{minipage}{0.94\linewidth}
    \small
    \textbf{#1}\par\smallskip
    \itshape #2
    \end{minipage}%
  }%
  \par\smallskip
}
\newcommand{\conditionbox}[4]{%
  \par\smallskip\noindent
  \fcolorbox{#2}{#3}{%
    \begin{minipage}{0.94\linewidth}
    \small
    \textbf{#1}\par\smallskip
    #4
    \end{minipage}%
  }%
  \par\smallskip
}

\usepackage{fontawesome}

\title{How AI Agents Discover Scientific Equations: From Hydrotope Rediscovery to New Water-Wave Amplitudes}

\author{
\And
Zihan Zhou \\
Department of Physics,\\
Princeton University,\\
Princeton, NJ 08540, USA \\
\texttt{zihanz@princeton.edu}
\And
Digvijay Wadekar \\
Center for Gravitational Physics,\\
University of Texas at Austin,\\
Austin, TX 78712, USA \\
\And
Matias Zaldarriaga \\
School of Natural Sciences, \\
Institute for Advanced Study, \\
1 Einstein Drive, Princeton, NJ 08540, USA
}

\begin{document}
\maketitle

%==============================================================================================
%==============================================================================================
\begin{abstract}
We study how AI agents discover and validate scientific formulas using a controlled case
study of the
hydrotope~\citep{Arkani-Hamed2026SurfaceWW}, a recently discovered geometric
formula that combines the different polynomial pieces of nonlinear
surface-wave scattering into one global expression. This problem is deceptively difficult: simple formulas can hold within
individual frequency regions, but the global result must identify their
boundaries and combine exponentially many potentially active terms. We reconstruct how the formula was originally discovered through human--agent collaboration
and analyze 18 single-prompt rediscovery runs under no hint and two forms of
human guidance: a false hint representing an incorrect prior and a true hint
representing domain-informed insight.
Only four recover the formula across all kinematic chambers (i.e., regions in which
a single polynomial form applies), while most unsuccessful
runs find correct chamber polynomials but fail to combine them or test their full
domain. Conventional and LLM-assisted symbolic regression and standard
machine-learning regressors likewise fail to recover the global formula in our experiments. Guided by these failure modes, we test a
PI$+$two-student workflow in which a coordinating lead agent assigns complementary
analytic and numerical tasks to two research agents and independently evaluates
their results. The PI$+$two-student team successfully
rediscovers the complete hydrotope formula, while the same workflow applied to the
harder three negative wavenumber problem discovers a new independent verified
analytic expression for the six-point amplitude $A_6$. \href{https://github.com/ZihanZhou26/hydrotope_benchmark}{\faGithub}
\end{abstract}

%==============================================================================================
%==============================================================================================

%==============================================================================================
%==============================================================================================
\section{Introduction}

General-purpose AI systems are rapidly becoming useful tools for scientific
research. Automated scientific systems have long connected hypothesis
generation to physical experiments, and modern autonomous laboratories add
active learning and robotic execution
\citep{King2004RobotScientist,King2009AutomationScience,Burger2020MobileChemist,
Szymanski2023ALab}. Tool-using language agents can interleave reasoning with
external actions, select application-programming interfaces, edit code, and
execute tests
\citep{Yao2022ReAct,Schick2023Toolformer,Patil2023Gorilla,Yang2024SWEAgent}.
They have already planned computational and laboratory tasks in chemistry,
while broader autonomous-research systems connect literature review, code
execution, experimentation, and scientific writing
\citep{Bran2024ChemCrow,Boiko2023AutonomousResearch,Lu2024AIScientist,
Schmidgall2025AgentLaboratory,VillaescusaNavarro2025Denario}. By combining
language-based reasoning with code execution, symbolic computation, and
numerical experimentation, recent systems have discovered new mathematical
constructions and algorithms \citep{funsearch,alphaevolve,Wang2025AutomatedAD}.

More recently, general-purpose models have begun to contribute directly to
open problems in mathematics and theoretical physics. An OpenAI model
autonomously disproved Erdős's longstanding conjecture on the planar
unit-distance problem \citep{openai2026unitdistance}; the report credits Claude
Fable with finding an explicit three-dimensional counterexample to the Jacobian conjecture
\citep{ramos2026jacobian}; and OpenAI reported ten further advances that
resolve or substantially advance longstanding problems across mathematics and
theoretical computer science \citep{openai2026tenadvances}. In scattering
amplitudes, GPT-5.2 conjectured a formula valid for any number of waves for half-collinear
single-minus gluon tree amplitudes, which another model subsequently proved
and researchers verified analytically \citep{guevara2026singleminus}. These
developments make it possible to study how AI systems contribute to scientific
discoveries, rather than judging them only by their final answers.

Despite this progress, we still know little about how an AI system reaches a
scientific result. Existing benchmarks test agents in interactive settings and
scientific experiments, but they also show that agents struggle with
many-step plans, data analysis, and completing an entire research task
\citep{Liu2023AgentBench,Huang2023MLAgentBench,Chen2024ScienceAgentBench,
Gu2024BLADE}. Some newer evaluations therefore record partial progress and the
full sequence of actions, not just whether the final answer is correct
\citep{Ma2024AgentBoard,Lu2025AgentRewardBench}. This record can show how an
agent proposes and changes a formula, what useful intermediate results it
finds, how it responds to a counterexample, and when it stops checking and outputs a solution. These details
matter because an agent may derive several correct formulas for separate cases
without combining them, or may propose one short formula without testing all
the edge-cases where it is supposed to work. We need a problem for which the
intermediate calculations, final answer, and tests can all be checked.

The recent discovery of the \emph{hydrotope} provides such an opportunity
\citep{Arkani-Hamed2026SurfaceWW}. Geometrically, the hydrotope is a slice
through a frequency-dependent box; its volume encodes one formula for the
scattering amplitude across all frequency regions.

The underlying problem arises in one-dimensional
deep-water surface-wave scattering. Linearizing the fluid equations makes
infinitesimal waves obey superposition, but finite-amplitude
waves satisfy nonlinear free-surface boundary conditions: products of the
surface displacement and fluid velocity couple different frequencies, allowing
the waves to exchange energy and scatter. The Berends--Giele (BG) recursion, an
algorithm that builds an $n$-point amplitude from lower-point interactions, can
generate exact interaction strengths at chosen frequencies. The recursion gives the answer at a
chosen point, but does not reveal where the formula changes or how the
different formulas fit together. The scattering amplitude is not described by the same polynomial formula everywhere in
frequency space. Within one
region, one polynomial formula applies; after crossing a boundary into a
neighboring region, a different polynomial takes over. We call these regions
\emph{frequency chambers} and their boundaries \emph{chamber boundaries}. The
familiar function $|x-y|$ gives the simplest example: it equals $x-y$ in the
region $x>y$ and $y-x$ in the region $y>x$, with the boundary at $x=y$. The wave
amplitude contains many analogous thresholds involving combinations of the
input frequencies, so its frequency space is divided into many such regions.
The challenge is not merely to fit the polynomial within one chamber. A
complete solution must find every relevant chamber boundary and combine the
polynomials on all sides of those walls.

This combination of exact query access and chamber-dependent formulas makes the
problem a useful case study of scientific discovery. Human researchers and an
AI agent originally obtained the all-$n$ formula through an extended
interaction, beginning with a simple expression
valid in one chamber and ending with a single formula valid across all
chambers. The researchers preserved the prompt sequence and computational
evidence, so this episode allows us to reconstruct an actual human--agent
discovery process and repeat the problem under fixed conditions.

The agent reached the formula by repeatedly proposing an equation, checking it
with BG recursion, and revising it after a failed test
\citep{Arkani-Hamed2026SurfaceWW}. A failure could reveal a missing chamber
wall or show that an equation worked in fewer chambers than expected. Related
language-agent methods also use outside feedback to revise an answer
\citep{madaan2023selfrefine,Shinn2023Reflexion,Gou2024CRITIC,Wu2024ProCo}.

\begin{figure}[H]
\centering
\includegraphics[width=\linewidth]{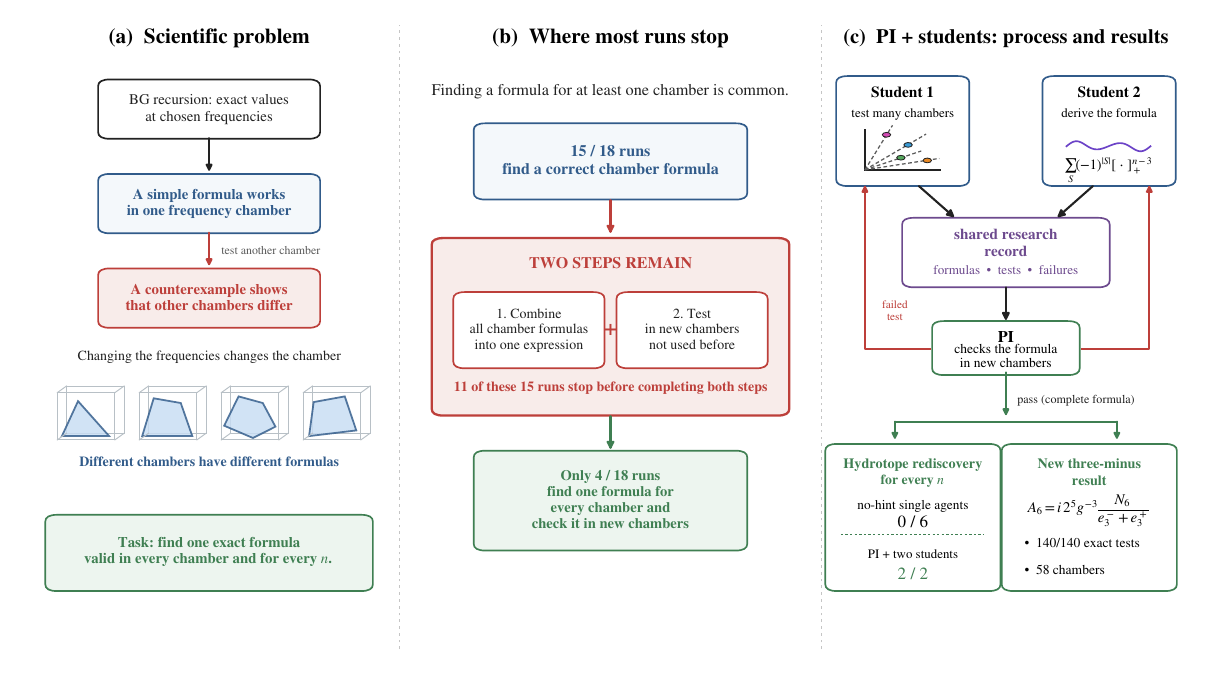}
\caption{This paper asks how AI agents can infer an analytic scattering
formula from numerical values provided at different input frequencies. (a) Berends--Giele (BG)
recursion numerically calculates the exact answer at any allowed point but does not reveal the analytic formula by itself. A simple formula
can work in one frequency chamber and fail in another, so the task is to find
one expression valid in every chamber and for an arbitrary number of waves. (b)
Fifteen of 18 single-agent runs find a correct formula in at least one chamber,
but only four combine the chamber formulas and test the result in new
chambers. (c) Two students try different approaches and write their
formulas, tests, and failures in a shared record. A PI combines their
work and checks it on new examples. Neither of six single-agent instances finds
the complete all-$n$ hydrotope formula, whereas both PI$+$student agentic teams do. The same team
also produces a new six-point three-minus formula that passes 140 exact tests
in 58 chambers.}
\label{fig:paper-overview}
\end{figure}

To separate this interactive search from equation fitting on a fixed dataset,
we also compare the agent-discovered expression with conventional symbolic
regression, LLM-assisted symbolic regression, and standard machine-learning
methods. This comparison asks whether these methods can recover or approximate
the amplitude directly from the input frequencies; it is distinct from the
repeated discovery runs described later.

Several recent AI-assisted physics results share one useful feature: although
the desired formula is unknown, a computer program can check any proposed
formula. Here BG recursion gives exact amplitude values at chosen frequencies,
so a failed proposal comes with a precise counterexample. FunSearch and
AlphaEvolve similarly use executable checks while searching for mathematical
constructions and algorithms \citep{funsearch,alphaevolve}. In a recent study
of single-minus gluon amplitudes, GPT-5.2 Pro proposed an all-$n$ formula,
another model proved it, and the authors checked it by hand with BG recursion
\citep{guevara2026singleminus}. Other studies used numerical simulations to
check formulas for gravitational waves, or a Monte Carlo calculation to check
a perturbative-QCD result
\citep{Islam2026InterpretableSurrogates,Islam2026UnifiedRemnant,
Schwartz2026CParameter}. Such exact calculations or accurate simulations make
wrong proposals easy to reject. They do not, however, tell the agent what
formula to try. Nor can a finite set of successful tests prove that a formula
works everywhere. Finding the formula and testing all the cases in its stated
range remain separate and difficult tasks.

We first reconstruct the prompt sequence that led to the hydrotope, then
analyze 18 single-prompt rediscovery runs under no hint, a false hint, and a
true hint. The false hint represents an incorrect human suggestion, whereas the
true hint supplies useful mathematical information without revealing the
answer. Only four runs recover the complete all-chamber formula. Many of the
others still find correct chamber polynomials and essential parts of the final
structure, but they fail to combine those pieces or accept a candidate without
testing it across all chambers. The unsuccessful runs therefore locate the
bottleneck between finding a local pattern and completing a global formula.

We use that diagnosis to design a team that separates formula construction
from checking. Two students pursue complementary questions, while a PI combines
their results, identifies missing cases, and tests formulas outside the
chambers used to derive them. In two no-hint experiments without access to the
published answer, both teams automatically rediscover the complete hydrotope
formula. Applied to the harder three-minus sector, the same workflow produces a
new, independently checked expression for the six-point amplitude $A_6$.
The teams use more computation than the single agents, so these runs do not
isolate which part of the team causes the improvement. Extensive exact tests
support the new $A_6$ expression, but we do not provide an analytic proof, and
its numerator for an arbitrary number of waves remains open.
Figure~\ref{fig:paper-overview} summarizes the problem, the team, and the two
scientific results.

\section{Surface-water-wave scattering and the discovery of the hydrotope}

We first introduce the physical problem and the few mathematical terms needed
to follow the discovery. We then state the hydrotope formula and reconstruct
the prompts that turned a formula valid in one frequency chamber into a
complete formula.

\subsection{Surface water wave scattering}

On deep water, a wave with spatial frequency, or \emph{wavenumber}, $k$ has
temporal frequency fixed by
$\omega^2=g|k|$, where $g$ is the gravitational acceleration. A single wave
propagates freely, but the nonlinear free surface couples several waves and
allows them to exchange energy. The coefficient $A_n$, called the $n$-point
amplitude, measures the strength of an interaction among $n$ waves; such coefficients are the microscopic input to
wave-turbulence descriptions of a wind-driven sea \citep{Newell2011WaveTurbulence}.

The physical process consists of $n-1$ incoming linear waves and one outgoing
nonlinear wave. For bookkeeping, we use the ``all-incoming'' convention and
assign each wave a signed frequency $\omega_i$ and a one-dimensional wavenumber
$k_i=\sigma_i\omega_i^2/g$. Here $\sigma_i\in\{-1,+1\}$ records whether the
wavenumber points left or right. A physical resonant configuration obeys the
free-wave dispersion relation for every wave while conserving total energy and
momentum. In scattering terminology, such a configuration is \emph{on shell}:
\begin{equation}
\label{eq:energy_momentum_conserve}
\sum_{i=1}^n\omega_i=0,\qquad
\sum_{i=1}^n\sigma_i\omega_i^2=0 ~.
\end{equation}
This is analogous to specifying a collision in which every participant obeys
its own energy--momentum relation and the total energy and momentum balance;
an arbitrary list of frequencies is therefore not an allowed query point.

We organize configurations by the signs $\sigma_i$. The first class with a
nonzero interaction has two negative wavenumbers and $n-2$ positive ones,
\begin{equation}
\sigma=(-1,-1,+1,\ldots,+1),
\end{equation}
which we call the \textbf{two-minus sector}. The name counts wavenumber
directions; it does not mean that two physical waves are absent or have
negative energy.
For example, at five points the sign pattern
$(-1,-1,+1,+1,+1)$ contains two waves with negative wavenumber and three
with positive wavenumber.

Earlier analytic work established important results at four and five points.
Zakharov gave a systematic energy-based description of surface waves
\citep{Zakharov1968}, and
Dyachenko and Zakharov showed that the energy-conserving four-wave interaction vanishes
\citep{DyachenkoZakharov1994}. The five-wave interaction does not vanish, but
Lvov obtained it only by summing 81 perturbative contributions and treating
many frequency chambers separately; the final expressions were nevertheless
remarkably simple \citep{DyachenkoLvovZakharov1995,Lvov1997}. No organizing
formula valid for arbitrary $n$ was then known.

The BG recursion offers a complementary computational tool. It builds an
$n$-point amplitude from lower-point sub-amplitudes and can evaluate $A_n$
exactly at a chosen on-shell point without enumerating every diagram by hand.
Recursive constructions are a standard way to organize scattering amplitudes
\citep{Berends1988Recursive,Britto2005BCFW}.
For a computer-science reader, it is similar to dynamic programming: the
recursion combines previously computed lower-point subproblems to evaluate a
larger one.
Its output, however, is a value rather than an explanatory formula. The
discovery task is therefore to convert exact query access into one analytic
expression for the two-minus amplitude, valid for every $n\geq4$ and throughout
the allowed frequency space \citep{Arkani-Hamed2026SurfaceWW,Cao:2026bvk}.

In the $g=1$ units used by the benchmark, that formula is
\begin{equation}
A_n =
i\,2^{n-1}\omega_1\omega_2
\sum_{S\subseteq\{3,\ldots,n\}}(-1)^{|S|}
\left(\beta^2-\sum_{j\in S}\omega_j^2\right)_+^{n-3},
\qquad
\beta=\min(|\omega_1|,|\omega_2|),
\label{eq:key}
\end{equation}
where $(x)_+=\max(x,0)$. Although short, it contains one term for every subset
$S$ of the waves with positive wavenumber, and $|S|$ is the number of waves in that
subset. For example, at $n=5$, $S=\varnothing$ gives the term $\beta^4$,
$S=\{3\}$ gives $-(\beta^2-\omega_3^2)_+^2$, and $S=\{3,4\}$ gives
$+(\beta^2-\omega_3^2-\omega_4^2)_+^2$. The sum runs over all eight subsets
of $\{3,4,5\}$, with the sign alternating as waves are added to $S$.

A term contributes only when its
squared-frequency sum lies below $\beta^2$. The equalities
$\sum_{j\in S}\omega_j^2=\beta^2$ are \emph{chamber boundaries}: crossing one turns
a term on or off. Between boundaries, the same terms remain active and the amplitude
is one polynomial; the $\max(x,0)$ notation combines all such chamber
polynomials into one expression.
For example, if $\beta^2=10$, a subset with squared-frequency sum $7$
contributes through $(10-7)_+=3$, whereas a subset with sum $12$ contributes
zero; the value $10$ is the chamber boundary between these cases.

\begin{figure}[H]
\centering
\begin{tikzpicture}[x=0.72cm,y=0.72cm,every node/.style={font=\scriptsize}]
  \node[font=\small\bfseries] at (1.85,2.15) {$n=5$: polygon chambers};
  \begin{scope}[shift={(0,0)}]
    \fill[waterblue!20] (0,0) -- (1.0,0) -- (0.5,0.88) -- cycle;
    \draw[wateredge,thick] (0,0) -- (1.0,0) -- (0.5,0.88) -- cycle;
    \node at (0.5,-0.42) {triangle};
  \end{scope}
  \begin{scope}[shift={(1.85,0)}]
    \fill[waterblue!20] (0,0) -- (1.2,0) -- (0.9,0.8) -- (0.3,0.8) -- cycle;
    \draw[wateredge,thick] (0,0) -- (1.2,0) -- (0.9,0.8) -- (0.3,0.8) -- cycle;
    \node at (0.6,-0.42) {trapezoid};
  \end{scope}
  \begin{scope}[shift={(3.95,0)}]
    \fill[waterblue!20] (0,0.15) -- (0.35,0.85) -- (1.25,0.85) -- (1.55,0.35) -- (1.05,-0.15) -- cycle;
    \draw[wateredge,thick] (0,0.15) -- (0.35,0.85) -- (1.25,0.85) -- (1.55,0.35) -- (1.05,-0.15) -- cycle;
    \node at (0.78,-0.42) {pentagon};
  \end{scope}
  \begin{scope}[shift={(6.35,0)}]
    \fill[waterblue!20] (0.25,0) -- (1.35,0) -- (1.1,0.85) -- (0,0.85) -- cycle;
    \draw[wateredge,thick] (0.25,0) -- (1.35,0) -- (1.1,0.85) -- (0,0.85) -- cycle;
    \node at (0.68,-0.42) {parallelogram};
  \end{scope}
  \begin{scope}[shift={(8.65,0)}]
    \fill[forbidred!10] (0.25,0) -- (1.0,0) -- (1.35,0.55) -- (1.0,1.1) -- (0.25,1.1) -- (-0.1,0.55) -- cycle;
    \draw[forbidred,thick,dashed] (0.25,0) -- (1.0,0) -- (1.35,0.55) -- (1.0,1.1) -- (0.25,1.1) -- (-0.1,0.55) -- cycle;
    \node at (0.62,-0.42) {hexagon};
    \node[forbidred] at (0.62,-0.78) {forbidden};
  \end{scope}

  \node[font=\small\bfseries] at (1.95,-2.15) {$n=6$: three-dimensional solids};
  \begin{scope}[shift={(0,-4.05)}]
    \fill[waterblue!18] (0.0,0.0) -- (1.2,0.15) -- (0.62,0.95) -- cycle;
    \fill[waterblue!28] (0.0,0.0) -- (0.62,0.95) -- (0.45,-0.78) -- cycle;
    \fill[waterblue!12] (1.2,0.15) -- (0.62,0.95) -- (0.45,-0.78) -- cycle;
    \draw[wateredge,thick] (0.0,0.0) -- (1.2,0.15) -- (0.62,0.95) -- cycle;
    \draw[wateredge,thick] (0.0,0.0) -- (0.45,-0.78) -- (1.2,0.15);
    \draw[wateredge,thick] (0.45,-0.78) -- (0.62,0.95);
    \node at (0.62,-1.25) {tetrahedron};
  \end{scope}
  \begin{scope}[shift={(2.35,-4.05)}]
    \fill[waterblue!18] (0,0) -- (1.3,0) -- (1.0,0.75) -- (0.25,0.8) -- cycle;
    \fill[waterblue!25] (0,0) -- (0.25,0.8) -- (0.45,-0.65) -- cycle;
    \fill[waterblue!12] (1.3,0) -- (1.0,0.75) -- (1.55,-0.55) -- cycle;
    \draw[wateredge,thick] (0,0) -- (1.3,0) -- (1.0,0.75) -- (0.25,0.8) -- cycle;
    \draw[wateredge,thick] (0,0) -- (0.45,-0.65) -- (1.55,-0.55) -- (1.3,0);
    \draw[wateredge,thick] (0.45,-0.65) -- (0.25,0.8);
    \draw[wateredge,thick] (1.55,-0.55) -- (1.0,0.75);
    \node at (0.78,-1.25) {prism};
  \end{scope}
  \begin{scope}[shift={(5.0,-4.05)}]
    \fill[waterblue!18] (0.25,0.0) -- (1.2,0.0) -- (1.55,0.55) -- (0.98,1.02) -- (0.25,0.92) -- (-0.12,0.42) -- cycle;
    \fill[waterblue!28] (0.25,0.0) -- (-0.12,0.42) -- (0.1,-0.52) -- cycle;
    \fill[waterblue!12] (1.2,0.0) -- (1.55,0.55) -- (1.42,-0.42) -- cycle;
    \draw[wateredge,thick] (0.25,0.0) -- (1.2,0.0) -- (1.55,0.55) -- (0.98,1.02) -- (0.25,0.92) -- (-0.12,0.42) -- cycle;
    \draw[wateredge,thick] (0.1,-0.52) -- (0.25,0.0);
    \draw[wateredge,thick] (0.1,-0.52) -- (-0.12,0.42);
    \draw[wateredge,thick] (0.1,-0.52) -- (1.42,-0.42) -- (1.2,0.0);
    \draw[wateredge,thick] (1.42,-0.42) -- (1.55,0.55);
    \node at (0.75,-1.25) {truncated};
  \end{scope}
  \node[anchor=west,align=left] at (7.15,-4.0) {14 ways the cut meets faces\\12 typical solids\\all from the same\\$\max(x,0)$ sum};
\end{tikzpicture}
\caption{Geometric interpretation of the two-minus hydrotope, the known
formula for the sector in which two waves have negative wavenumber. A
chamber-dependent scattering amplitude can be represented by a slice of a box
whose edge lengths are set by the wave frequencies. Crossing a chamber
boundary switches a term $(x)_+=\max(x,0)$ in Eq.~\ref{eq:key} on or off, so
the sum adds the active contributions and subtracts their overlaps. At $n=5$,
a plane cuts a three-dimensional box into four allowed polygon types, with the
conservation laws excluding the central hexagon; at $n=6$, the analogous cut
is in four dimensions and gives different three-dimensional solids in
different chambers. One formula built from $\max(x,0)$ therefore describes
every chamber-dependent cross-section.}
\label{fig:waterhedron-geometry}
\end{figure}
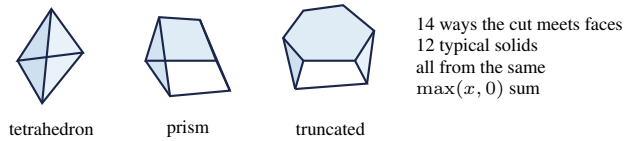

The formula also has a geometric interpretation. Define the \textbf{hydrotope}
\begin{equation}
\mathcal{W}_n=\left\{(t_3,\ldots,t_n):0\leq t_i\leq \omega_i^2,\
\sum_{i=3}^n t_i=\beta^2\right\},
\end{equation}
the part of a box that also satisfies the displayed sum. This flat cut changes
shape as the frequencies vary. Apart from a common factor, $A_n$ is its volume,
computed by adding individual pieces and subtracting their overlaps. For two
overlapping regions $A$ and $B$, the familiar identity
$\operatorname{vol}(A\cup B)=\operatorname{vol}(A)+\operatorname{vol}(B)
-\operatorname{vol}(A\cap B)$ is the simplest example of the same accounting
principle. The resulting shape is a polygon at $n=5$ and a three-dimensional
solid at $n=6$, so the many chamber polynomials are different slices of the same
object, as illustrated in Figure~\ref{fig:waterhedron-geometry}.

\subsection{Original discovery episode}

An earlier Claude Code session in the same water-wave project, a human-guided
research episode in which a user and the agent worked interactively, first
produced Eq.~\eqref{eq:key}. The preserved history records the user's prompts but
not the assistant's replies. Work on the two-minus sector began on March 13,
2026, after the researchers built and checked the Berends--Giele code. The first
discovery-relevant prompt fixed the two-minus convention and proposed a candidate
formula for one frequency region:

\promptbox{March 13, 2026, 22:03:23}{
can you test the 1D case numerically with all positive but two negative (to make
sure we are on the same convention. I choose the first and second leg to have
$\sigma$ negative) from four point all the way to eight or nine point. I guess
the expression would be proportionl to
$\omega_1 \omega_2 \min(\omega_1^2, \omega_2^2)^{n-3}$, the n the n-point
amplitude
}

The prompt carries genuine physical intuition. It requests a sweep over the
number of waves and supplies the one-term product
$\omega_1\omega_2\min(\omega_1^2,\omega_2^2)^{n-3}$, which many agents later
rediscover on their own. It also contains, in miniature, the central difficulty:
this expression is exact in one frequency chamber but is not the complete formula.

\promptbox{March 13, 2026, 22:45:58}{
try to relax the assumption, $\omega_2\ldots \omega_{n-1}>0$,
$|\omega_2|>|\omega_1|$ and guess the general formula
}

The second prompt turns a check in one chamber into a search for the general formula.
By relaxing the ordering and positivity assumptions under which the agent
obtained this formula, it pushes the search beyond the chamber in which the
agent fitted it, toward an expression that remains valid under arbitrary signs and
frequency orderings.

\promptbox{March 13, 2026, 23:04:14}{
For the failed case, try to guess the formula
}

The third prompt returns to the configuration where the one-term formula had failed and
asks for the corrected formula. From there the agent finds the
alternating subset sum in Eq.~\eqref{eq:key}, which adds the contributions
from different chambers and subtracts their overlaps, and checks it exactly at
higher $n$. Across the three prompts, the task changes in three steps: propose
a formula for one chamber, ask for a formula that works more broadly, and use
the failed case to find the complete formula. The rest of the paper asks
whether agents can complete these steps without those follow-up prompts.

\section{Experiments and comparison methods}
\label{sec:experimental-design}

We study the task in two ways. First, we repeat it with different prompts and
instructions for choosing test points. Second, we ask whether symbolic
regression and standard machine-learning methods can learn the formula from a
fixed table of BG results.

\subsection{Repeating the discovery task}
\label{sec:controlled-replay}

Because Eq.~\eqref{eq:key} was new and unavailable publicly when we performed
these experiments, the agents could not retrieve it from the
literature. They therefore had to discover the result from scratch. This allows
us to study the discovery process itself, rather than the ability of an agent
to retrieve a known formula.

In each run, the agent receives only a prompt and the Mathematica code
\texttt{OnShellBG.m}, which numerically evaluates the amplitudes using the
Berends--Giele recursion. We compare the three prompt packages summarized in
Table~\ref{tab:prompt-packages}; they vary both the suggested type of equation
and the recommended tests.

\begin{table}[H]
\caption{The repeated runs vary both the suggested type of equation and the
recommended tests. Appendix~\ref{app:prompt-packages} reproduces the
complete prompt descriptions.}
\label{tab:prompt-packages}
\centering
\small
\begin{tabularx}{\linewidth}{@{}lYY@{}}
\toprule
package & equation guidance & recommended tests \\
\midrule
false hint & one ratio of polynomials for all frequencies; no chamber split &
comparable frequencies; avoid separated scales and chamber boundaries \\
true hint & different polynomials in different chambers, with fixed scaling &
find the chamber boundaries and test on both sides \\
no hint & no proposed equation class &
test $n=4,5,6,7$, including widely separated frequency scales \\
\bottomrule
\end{tabularx}
\end{table}

The false-hint package represents an incorrect suggestion from a human and
tests whether exact counterexamples make the agent abandon it. The true-hint
package supplies useful mathematical information: it says that different
polynomials apply in different chambers, but does not give the chamber boundaries,
coefficients, or complete formula. The no-hint package suggests no type of
equation. Because the instructions for choosing test points also differ, we
compare the three complete prompt packages, not just the hints.

We perform 18 independent runs: one run for each of six agent configurations
under each of the three prompt conditions. The configurations are Claude Opus
4.8 max, Claude Opus 4.8 ultra, Codex 54 xhigh, Codex 55 xhigh, DeepSeek v4 pro,
and Fugu ultra. Section~\ref{sec:single-prompt-results} compares their
outcomes and analyzes the recorded steps that led to them. Because each
configuration appears only once per condition, we report the observed outcomes
rather than pass@$k$: estimating pass@$k$ would
require several independent runs for every
configuration and condition, which we did not run because the long,
tool-using searches incur substantial token usage and compute cost. The counts
therefore describe these 18 runs rather than estimate each configuration's
probability of success.

We provide the prompts, run histories, notes, checking code, and
benchmark datasets at
\url{https://github.com/ZihanZhou26/hydrotope_benchmark}. By releasing the
complete prompts and checking code, we invite the community to test other
single agents and teams of agents and to compare their success rates,
run histories and resource use under explicitly reported token and compute
budgets.

\subsection{Symbolic regression and machine-learning comparisons}
\label{sec:symbolic-regression}

A natural alternative to the agents' propose--test--revise search is
\emph{symbolic regression} (SR). It searches a table of numerical inputs and
outputs for an equation, favoring equations that are both accurate and short.
Unlike an ordinary regression model, its result is meant to be read by a
person. Related scientific machine-learning methods also use known physics to
guide the search \citep{Carleo2019MLPhysicalSciences,Karniadakis2021PIML}.
One type of SR selects terms from a list supplied in advance
\citep{Brunton2015SINDy}. Another builds equations from the allowed variables
and operations, keeps the better equations, and repeatedly changes or combines
them
\citep{Schmidt2009NaturalLaws,Bongard2007ReverseEngineering,
Burlacu2020OperonCA,LaCava2021SRBench,Makke2022SRReview}.
Other versions use known symmetries, separate independent parts of the
problem, or learn useful variables before searching
\citep{Martius2016Equations,Udrescu2020AIFeynman,
Cranmer2020SymbolicModels,Biggio2021NeuralSR,Kamienny2022TransformerSR}. These
methods can test many equations in parallel. Genetic symbolic regression, for
example, returns candidates with different balances between accuracy and
simplicity, although the result can depend strongly on the supplied variables,
allowed operations, training points, and search time
\citep{Udrescu2020AIFeynman,Cranmer2023InterpretableML}. Across these
approaches, the search usually responds to overall error rather than asking
what one failed example means physically.

A language-model agent can use the same numbers in a different way. It can
define a quantity suggested by the physics, choose a chamber that separates
two possible formulas, ask for a calculation likely to disprove its current
formula, and revise either the formula or the chambers where it claims the
formula works. It can also change the kind of formula it is searching for,
rather than combining operations from one fixed list. This freedom can reveal
a pattern that appears rarely in a fixed dataset, but it also makes the result
more dependent on the prompt and the agent's choices. An agent can follow an
unproductive idea or accept a formula after too few tests, so every proposal
must be checked.

Hybrid SR methods attempt to combine these strengths. LaSR uses an LLM to
extract and evolve concepts within a population-based search, whereas LLM-SR
uses an LLM to propose equation programs and then optimizes and selects them
against data
\citep{Grayeli2024SymbolicRW,Shojaee2024LLMSRSE}. Related neural and hybrid
methods generate expressions with reinforcement learning or transformers, or
seed genetic search with neural proposals
\citep{Petersen2019DeepSR,Valipour2021SymbolicGPT,Mundhenk2021NeuralGuidedGP}.

More broadly, several recent methods combine LLMs, symbolic regression, and
iterative scientific reasoning
\citep{Yang2026ThinkLA,su2026strideselfreflectiveagentframework,xia2025sr,
wang2025drsr,guo2025sr,pang2026deliberateevolutionagenticreasoning,N_gele_2026}.
In a separate gravitational-wave problem, Islam et al. report a related
result: their agent-built formula, checked by simulations, was more accurate
than the symbolic-regression and conventional ML methods and was also short
enough to interpret physically
\citep{Islam2026InterpretableSurrogates}.
Other scientific methods similarly combine learned variables, short equations,
physical knowledge, or language-model criticism to go beyond unrestricted
curve fitting
\citep{Wu2018AIPhysicist,Champion2019Coordinates,Chen2021StateVariables,
Cornelio2021AIDescartes,Li2024AutomatedModelDiscovery}.

For this comparison, we use the Berends--Giele recursion as a data generator
and benchmark three conventional
SR systems (PySR, gplearn, and Operon), these two LLM-assisted systems, and
four numerical regression methods: linear regression, random forest, a
multilayer perceptron neural network, and a Kolmogorov--Arnold network (KAN).
Unlike a multilayer perceptron with fixed activation functions on its nodes, a
KAN learns univariate spline functions on its edges \citep{Liu2024KAN}. The ML
methods test whether they can predict the amplitude at new points drawn from the same distribution; the
SR methods additionally test whether the data reveal an explicit equation.

We consider both $n=5$, the first nontrivial two-minus case, and $n=7$, where
the complete expression is substantially larger. We remove the common factor
in Eq.~\eqref{eq:key} and use the rescaled amplitude
$\Phi_n=A_n/(2^{n-1}i\omega_1\omega_2)$ as the prediction target.
Writing $\beta^2=\min(\omega_1^2,\omega_2^2)$, the genuine multi-chamber
structure begins at $n=5$. The normalized form of Eq.~\eqref{eq:key} then expands to
\begin{align}
\Phi_5={}&\beta^4
-[\beta^2-\omega_3^2]_+^2
-[\beta^2-\omega_4^2]_+^2
-[\beta^2-\omega_5^2]_+^2\nonumber\\
&+[\beta^2-\omega_3^2-\omega_4^2]_+^2
+[\beta^2-\omega_3^2-\omega_5^2]_+^2\nonumber\\
&+[\beta^2-\omega_4^2-\omega_5^2]_+^2
-[\beta^2-\omega_3^2-\omega_4^2-\omega_5^2]_+^2.
\label{eq:sr-n5-expanded}
\end{align}
Even this first nontrivial case contains eight terms that can switch on or off and
several possible chamber boundaries. Adding another wave with positive wavenumber doubles
the number of terms: $\Phi_7$ contains 32 terms of the form
$[x]_+^4$. A candidate can therefore achieve a low average regression error
while omitting a term that activates only in a sparsely sampled chamber.
Each $\max(x,0)$ term switches on when its argument becomes positive, so a
dataset containing varied frequencies crosses several frequency chambers.

For each value of $n$, every fitted method, including conventional ML,
conventional and LLM-assisted SR, and the fixed-data agent in
Appendix~\ref{app:sr-diagnostics}, uses the same fixed train--test split. The
PI--student teams instead choose new BG calculations as they work rather than
train on this fixed table; we evaluate their final formula on the
same held-out set used in the plotted comparison. For a candidate prediction
$\widehat\Phi_n$, we define the held-out
mean absolute error as
\begin{equation}
\mathrm{MAE}_n
=\frac{1}{N_{\rm test}}\sum_{a=1}^{N_{\rm test}}
\left|\widehat\Phi_n(\boldsymbol\omega^{(a)})
-\Phi_n(\boldsymbol\omega^{(a)})\right|,
\label{eq:mae}
\end{equation}
where $N_{\rm test}$ is the number of held-out points and $\Phi_n$ is the exact
recursion-generated target. All fitted methods receive only the frequencies,
not the precomputed building blocks
$[x]_+$ in Eq.~\eqref{eq:key}. The SR methods may use arithmetic and
operations such as minimum and maximum, but we do not give them the
target formula, its frequency chambers, the frequency ordering, or how to divide
the formula into terms.

Increasing the amount of training data, especially in
rare frequency chambers, would likely improve the predictive accuracy of the
ML methods. However, the present ML methods use the full fixed training set,
whereas the successful discovery agents used fewer than 100 points to
construct their formulas. We also retrained the ML models on exactly the points
selected by the agents during discovery to test whether the choice of points
explained the difference; their
held-out predictive performance did not improve. More data alone, however, would not establish that a method had
recovered the exact complete formula rather than a more accurate approximation.

\begin{figure}[H]
\centering
\includegraphics[width=0.98\linewidth]{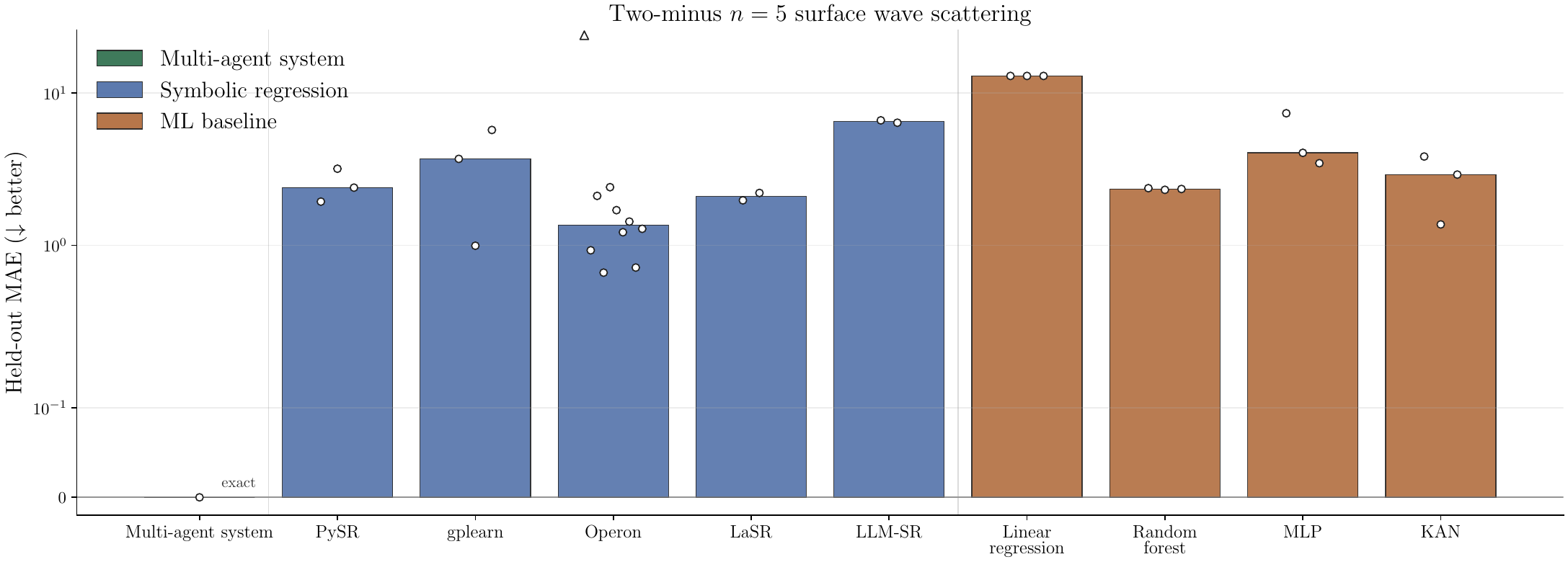}
\caption{As a benchmark to compare the performance of the agentic method, we test symbolic regression and classical ML methods (see section~\ref{sec:symbolic-regression}) to see if they can learn the relationship predicted by the exact (i.e., zero error) five-point hydrotope formula discovered by the multi-agent system (see Eq.~\eqref{eq:sr-n5-expanded}). We use the same training dataset used in the agentic runs: wave frequencies as input and corresponding output amplitude from the recursion (BG) relation. Bars show median
held-out mean absolute error (MAE), open markers show individual runs, and the
triangle marks an off-scale run retained in the benchmark table. Operon has
the lowest fitted median MAE (1.36), but no conventional, LLM-assisted, or prediction method recovers exact formula.}
\label{fig:symbolic-regression-n5}
\end{figure}

Figure~\ref{fig:symbolic-regression-n5} compares conventional and LLM-assisted
symbolic regression, standard machine-learning regressors, and the research
team on the five-point task. Bars show the median held-out MAE across runs,
and open markers show individual runs. Operon has the lowest median MAE among
the fitted methods, while random forest is competitive with several
equation-search systems. KAN reaches median MAE $2.91$ at $n=5$ and $927.04$
at $n=7$; it is the lowest-error numerical ML method at $n=7$, but it neither
reaches zero error nor returns a closed-form equation. Every fitted method
retains nonzero error, and no equation-search system recovers the complete
formula. The PI--student team is not a symbolic-regression method;
its zero-error entry evaluates the complete formula that it rediscovers.
Section~\ref{sec:multiagent} presents that team and its discovery results.
The corresponding $n=7$ test appears in
Figure~\ref{fig:symbolic-regression-n7} of Appendix~\ref{app:sr-n7}.

This comparison separates approximate prediction from formula discovery: a low
average error can come from a formula that misses chamber boundaries, $\max(x,0)$
terms that switch on, or entire frequency chambers. This distinction motivates
scientific-ML methods that produce explicit equations rather than treating
predictive accuracy alone as the endpoint
\citep{Rudin2018Interpretable,Cranmer2020SymbolicModels,
Cranmer2023InterpretableML,Bartlett2025SRPhysicalSciences}.

Additional experiments test two harder settings. At $n=7$, the target grows
from 8 to 32 $\max(x,0)$ terms; at $n=5$, an expanded search allows separate
formulas for different regions. No fitted method is exact: allowing
separate formulas can improve average prediction without recovering the
slanted chamber boundaries or the short alternating subset sum. Figure~\ref{fig:sr-appendix}
and Appendix~\ref{app:sr-diagnostics} give the additional systems, numerical
results, limits of the fixed train--test split, and the $n=7$ results.

The LLM-assisted methods can propose a wider variety of equations, but they
still select among them by average error on sampled data. Under our time and
compute limits, neither LaSR nor LLM-SR finds all the chambers or the complete
alternating subset sum. Replacing the earlier language model with a
stronger model does not change this result. An LLM inside an evolutionary or
program search does not automatically choose the counterexamples that expose a
missing chamber or change the kind of formula after such a failure.

Agents differ from these fitted methods in how they use a failed example.
Instead of minimizing only the average error, an agent can identify quantities
that should not change, choose a
useful chamber, find a wall where the proposed equation disagrees with BG,
test both sides of that wall, and revise the equation. In the original
discovery, this propose--test--revise cycle turned a one-term formula for one chamber
into the complete formula built from $\max(x,0)$. None of the conventional or
LLM-assisted methods tested here completes these steps.

The distinction between fixed-data fitting and interactive search also makes
the hydrotope a potential benchmark for comparing the two approaches.
Established SR test
collections compare equation-search methods with conventional
ML on synthetic and real-world regression problems, evaluate exact recovery,
and construct physics-derived datasets with realistic variable ranges
\citep{Orzechowski2018Benchmark,LaCava2021SRBench,Matsubara2022Rethinking}.
The hydrotope could complement these collections: it
is deterministic and exactly verifiable, but its inputs obey frequency
constraints, different polynomials apply in different chambers, and the number of
terms doubles each time another wave is added. Providing both a
fixed dataset and BG code that can calculate new points would support
comparisons between methods that fit a prescribed sample and interactive
methods that choose new evaluation points.

Readers should interpret this single comparison as a task-specific result
rather than a general ranking between language model agents and symbolic regression.
Performance depends on the allowed mathematical operations, search budget,
input variables, distribution of training points, and method-specific compute.
The comparison does not use equal compute and cannot show that any one design
choice caused the result. Appendix~\ref{app:sr-diagnostics} explains how we
selected formulas and the limits of the fixed train--test split.

\section{Single-prompt rediscovery: results and run histories}
\label{sec:single-prompt-results}

We analyze the 18 repeated discovery runs introduced in
Section~\ref{sec:controlled-replay}. We test each of six agent configurations
once with no hint, a false hint, and a true hint, and compare both their final
answers and the steps recorded during each run. After identifying the common
trajectory, we examine four no-hint runs that share the early steps but stop in
two different places: two produce an almost complete formula that still misses
cases, while two derive several chamber formulas but ultimately restrict their
claim to the principal chamber. Appendix~\ref{app:no-hint-details} gives the
formulas, test points, and counterexamples for each run.

We preserve the full record from every run in the accompanying repository. To
make these long records easier to read and compare, we rewrote each one as a chronological
account that organizes the hypotheses, commands, numerical
results, counterexamples, and final claims around the main changes in the
search. Each rewritten account is our interpretation of the corresponding
original log rather than a verbatim transcript. Accordingly, the process
claims below concern reasoning expressed in the preserved record; they do not
attribute unexpressed internal states to the agent. Unless stated otherwise,
first-person passages quoted from a run come from the rewritten account.
Appendix~\ref{app:thinking-logs} reproduces five complete rewritten records:
one successful true-hint run and the four no-hint runs examined in
Section~\ref{sec:four-no-hint}. The remaining 13 records are available in the
repository.

\subsection{Results and common steps}

We classify each run according to the strongest correct result contained in
its final answer. We use five mutually exclusive categories:

\begin{itemize}
    \item \outcome{complete}: the complete formula, valid in every kinematic
    chamber;
    \item \outcome{partial sum}: the correct alternating-sum structure, but
    with a missing term, an incorrect condition for including a term, or an
    untested case;
    \item \outcome{one chamber}: a correct all-$n$ formula in only one
    frequency chamber; a formula for all chambers at one fixed $n$ remains in
    this category until it is extended to arbitrary $n$;
    \item \outcome{hint rejected}: the agent correctly rejects the false single
    ratio-of-polynomials description but obtains no correct formula; and
    \item \outcome{incorrect}: the agent proposes the wrong type of formula or
    an incorrect equation, or produces no useful correct result.
\end{itemize}

Table~\ref{tab:outcome-counts} summarizes the results. Claude max, Claude
ultra, Codex 5.5, and Fugu produce all four \outcome{complete} formulas with
the true hint. The false-hint and no-hint runs produce no complete formula.
Thus, within these 18 runs, every complete result comes from a prompt that says
the answer is a different polynomial in different frequency regions.

\begin{table}[H]
\caption{Each of the three prompts is tested with six agent configurations.
``Complete'' means correct for every chamber and every $n$. ``Partial sum''
means that the agent finds the alternating-sum structure but misses a term,
an on/off condition, or a test case. ``One chamber'' means that the final
all-$n$ formula applies only in one chamber. All four complete results use the
true hint that says a different polynomial applies in each chamber.}
\label{tab:outcome-counts}
\centering
\small
\begin{tabular}{@{}lccccc@{}}
\toprule
condition & \outcome{complete} & \outcome{partial sum} & \outcome{one chamber} & \outcome{hint rejected} & \outcome{incorrect} \\
\midrule
false hint & 0 & 1 & 3 & 1 & 1 \\
true hint & 4 & 1 & 0 & 0 & 1 \\
no hint   & 0 & 2 & 4 & 0 & 0 \\
\midrule
total     & 4 & 4 & 7 & 1 & 2 \\
\bottomrule
\end{tabular}
\end{table}

Calling every run simply a success or a failure hides how far it gets. Of the
14 runs that do not reach the complete formula, only two are fully
\outcome{incorrect}. Eleven find at least one correct chamber polynomial, and
four of those also find the alternating-sum pattern but miss a term, an on/off
condition, or a broad enough test. The remaining run rejects the false
ratio-of-polynomials suggestion but does not find a replacement. Most runs
therefore find a useful part of the answer before stopping.

The outcome categories show how far each run gets; the recorded histories show
why runs exposed to different prompts begin to diverge.

\paragraph{How the three prompts affect the results}

The three prompts produce different results. With the true hint, four of the six runs recover the complete
formula, one obtains a partial sum, and one fails. The hint tells the agents to
look for a different polynomial in each chamber and therefore reduces the
range of formulas they need to consider. It still leaves the main calculation
to the agents: they must identify
the chamber boundaries produced by sums of frequencies, assemble the chamber polynomials into an
alternating subset sum, determine its overall factor, and verify that
the result holds outside the chambers used to build and test it.
The two incomplete runs fail at these later steps.

The false-hint runs give one partial sum, three one-chamber formulas, one
rejection, and one incorrect result. The difference comes from how the agents
respond when exact calculations contradict the suggested formula and test
points. Some agents abandon the
proposed ratio-of-polynomials form and reconstruct a formula with a different
polynomial in each chamber. Others recognize that this form fails but stop after rejecting it,
while still others continue searching within the wrong class of equations.
The false hint tests whether an agent can overrule confident but false
guidance, not merely whether it can fit a formula.

The no-hint runs show a different pattern. All six find something correct:
four stop at a one-chamber formula, while two reach an
almost correct alternating subset sum. These runs typically identify the principal-chamber formula,
find examples where it fails, and recognize that the amplitude changes across
chamber boundaries. Their main difficulty is not detecting chamber dependence, but
combining the formulas from different chambers, learning from failed attempts,
and testing the result in new chambers. The most common stopping point is
therefore the step from several correct chamber formulas to one formula valid
in every chamber. The comparison alone cannot tell us whether this would
remain the hardest step under different testing instructions.

\paragraph{Eight common steps}
We mark whether each run completes eight steps toward the known formula. Most
runs follow this order, although a run can partly complete a later step without
stating every earlier observation.

\begin{description}
    \item[\textbf{K1: Generate exact values.}]
    Construct a reliable BG recursion evaluator from \texttt{OnShellBG.m} and generate
    amplitude data.

    \item[\textbf{K2: Find properties that stay fixed.}]
    Identify three properties: $A_n$ is imaginary; rescaling every frequency
    by $c$ rescales it by $c^{2n-4}$; and it has sign-class symmetry, meaning
    that relabeling waves with the same wavenumber sign does not change it.

    \item[\textbf{K3: Principal-chamber formula.}]
    Recover the following one-term formula in the \emph{principal chamber}, where no
    correction terms are included. Here the smaller frequency among the two
    negative-wavenumber waves has the smallest magnitude:
    $\beta^2\leq\omega_j^2$ for every positive-wavenumber wave $j$. Consequently,
    every nonempty-subset $\max(x,0)$ term in Eq.~\eqref{eq:key} vanishes:
    \begin{equation}
    A_n\propto
    \omega_1\omega_2
    \min(\omega_1^2,\omega_2^2)^{n-3}.
    \end{equation}

    \item[\textbf{K4: Breakdown beyond the principal chamber.}]
    Find a concrete choice of frequencies for which the one-term formula does not
    reproduce the numerical amplitude.

    \item[\textbf{K5: Different formulas in different chambers.}]
    Interpret the counterexample as evidence that different chamber
    polynomials apply in different frequency chambers, rather than as numerical noise or a
    separate physical solution.

    \item[\textbf{K6: Find every boundary.}]
    Identify all chamber boundaries produced by sums of frequencies
    \begin{equation}
    \sum_{j\in S}\omega_j^2=\beta^2,
    \qquad
    \beta^2=\min(\omega_1^2,\omega_2^2),
    \end{equation}
    for arbitrary subsets $S$ of the positive-wavenumber waves.

    \item[\textbf{K7: Complete formula.}]
    Combine the chamber contributions into the complete alternating subset sum.

    \item[\textbf{K8: Test in new chambers.}]
    Test the formula in previously unsampled frequency chambers, near combined
    chamber boundaries involving sums of several frequencies, and at the exceptional
    $n=4$ boundary.
\end{description}

The first five steps cover local discovery: obtaining exact values, finding
stable properties, deriving one chamber formula, and recognizing that it fails
elsewhere. The last three steps require global completion: identifying every
boundary, combining the chamber contributions, and testing the result in
previously unsampled chambers.

To complement this stepwise classification with a numerical comparison, we
evaluate the most complete explicit
eight-point formula in each final answer. We remove the common factor
and use
\begin{equation}
\Phi_8(\boldsymbol\omega)
=\frac{A_8(\boldsymbol\omega)}{128i\omega_1\omega_2}
=\sum_{S\subseteq\{3,\ldots,8\}}(-1)^{|S|}
\left[\beta^2-\sum_{j\in S}\omega_j^2\right]_+^5,
\qquad \beta^2=\min(\omega_1^2,\omega_2^2).
\label{eq:checkpoint-phi8}
\end{equation}
The reported statistic is the held-out MAE defined in
Eq.~\eqref{eq:mae}, evaluated at $n=8$ with $\widehat\Phi_8$ set to the
strongest explicit formula in the final answer.
We generate the test set by drawing 1,000 sextuples
$(\omega_2,\ldots,\omega_7)$ from $\mathcal N(0,2.5^2)$ with seed 2025,
solving the two on-shell constraints for $\omega_1$ and $\omega_8$, and
reserving the last 250 rows. Some agents return a formula that applies only to
a restricted set of frequency configurations. We use the same expression to
predict all 250 evaluation points, including configurations outside that set,
so its MAE measures how far the partial formula is from an all-chamber result.
A dash marks final answers that stop short of an evaluable closed $n=8$
formula. This test is generated separately from the
five-point split used in Figure~\ref{fig:symbolic-regression-n5}.

\begin{figure}[H]
\centering
\includegraphics[width=0.98\linewidth]{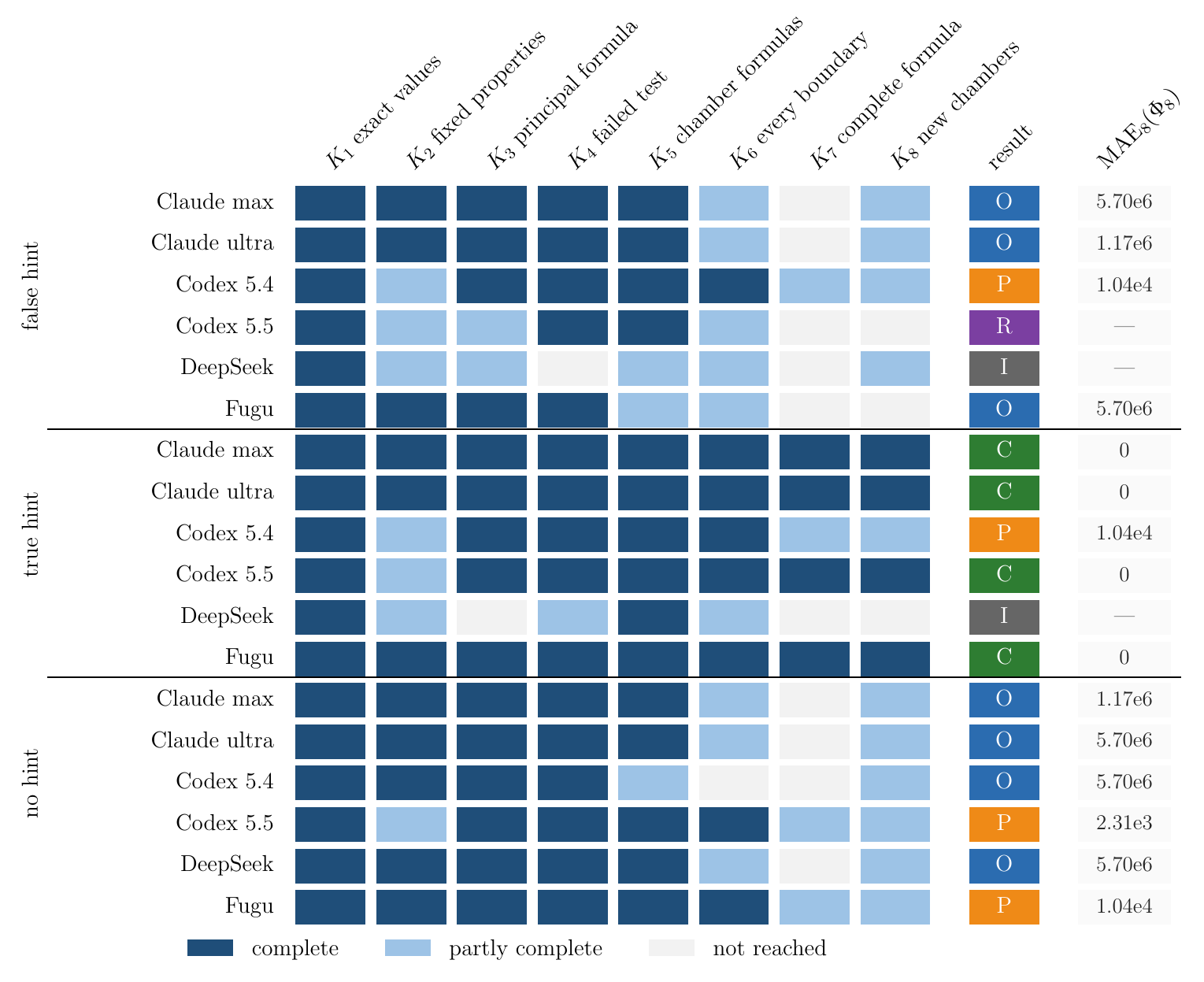}
\caption{The rows show eight key steps in order of completeness of the solution. These range from sampling
exact values from the BG recursion (K1) to testing and verifying the output analytic formula holds in new chambers (K8). Dark, light, and
pale cells mean complete, partial, and not reached. The result column uses
\textcolor{globalgreen}{\textbf{C}} for a complete formula,
\textcolor{nearorange}{\textbf{P}} for a partial sum,
\textcolor{localblue}{\textbf{O}} for a one-chamber formula,
\textcolor{rejectpurple}{\textbf{R}} for a rejected false hint, and
\textcolor{badgray}{\textbf{I}} for an incorrect result. The final column gives
the mean absolute error (MAE) of the run's $n=8$ formula on a separate set of 250
test points; a dash means that the final answer contains no formula that can be
evaluated at $n=8$. The largest difference among runs appears at K6--K7, where
they must find every boundary and combine the chamber terms.}
\label{fig:checkpoint-raster}
\end{figure}

Figure~\ref{fig:checkpoint-raster} shows that most runs complete the early
computational and mathematical steps. All 18 runs construct an evaluator at
K1. Eleven explicitly state all three fixed properties at K2, with sign-class
symmetry being the most frequently omitted. This omission does not appear to
prevent later progress. Fifteen runs recover the one-term principal-chamber formula at K3,
16 find a configuration in which that formula fails at K4, and 15 recognize
that the amplitude changes between chambers at K5.

The last column of Figure~\ref{fig:checkpoint-raster} compares the most complete
$n=8$ formula from each run with the exact amplitude on 250 test points. An MAE
of zero means that the formula matches every test point; a larger MAE means a
larger average error. The four runs that find the complete formula have zero
MAE. Among the incomplete results, the no-hint Codex~5.5 formula is the closest,
with MAE $2.31\times10^3$. An incomplete subset formula that leaves out the
wave with positive wavenumber fixed by the conservation equations has MAE
$1.04\times10^4$. A single-term formula valid only in the simplest frequency
region has MAE $1.17\times10^6$ when it treats the two negative-wavenumber waves
symmetrically and $5.70\times10^6$ when it always uses $\omega_2$. The false-hint Claude-ultra
run is labeled \outcome{one chamber} because it derives the complete formula only at
$n=5$; its $n=8$ score therefore comes from its simpler all-$n$ formula, which
applies in one chamber. These MAE values compare the accuracy of the incomplete
formulas. A separate chamber-by-chamber check is still required because an
average can hide errors that occur at only a few test points.

The largest difference among runs appears at K6 and K7. At K6, eight runs identify the complete
family of chamber boundaries associated with arbitrary subsets of the plus waves. Nine
others find only simpler chamber boundaries, such as single-wave thresholds or
orderings of the individual $|\omega_i|$, and one does not reach this stage.
At K7, only four runs combine the chamber contributions into the complete
alternating subset sum; these are exactly the four \outcome{complete} runs
under the true-hint condition. Four additional runs partially reach K7: they
find the correct alternating-sum pattern but use a boundary condition that is
valid only in their sampled chambers or include too few subsets.

The runs also differ in how broadly they test. Five runs test at least one
formula outside the frequency chambers in which the agents derived it, but only
four do so for an all-$n$ candidate. Ten runs
test only the comparatively simple configurations generated by the standard
sampling routine. The remaining three runs produce no formula suitable for
such tests. The overall pattern is therefore clear: most
agents can identify the formula within a restricted chamber and recognize
that the answer changes between chambers. Far fewer determine the complete set of chamber
walls, combine the chamber contributions into one formula, and verify that
the result holds throughout the two-minus sector.

\paragraph{Why runs stop before finding the complete formula}

The run histories show three common reasons:

\begin{enumerate}
    \item \textbf{A failed combination attempt leads back to a restricted answer.}
    The run establishes the principal-chamber formula, finds a counterexample, and
    computes additional chamber formulas. It then tries a formula based on the
    order of the frequencies, a universal absolute-value expression, or another
    way to combine the regions. The proposed formula misses boundaries that
    depend on sums of several frequencies and produces further
    counterexamples or a computer-algebra calculation that does not finish. The run then reports only
    the principal-chamber formula instead of using those failures to construct
    the alternating sum over subsets.

    \item \textbf{The formula is tested only on familiar configurations.}
    The run proposes an alternating subset sum, but tests it on
    the same nested family of frequency chambers used to construct it.
    Those tests do not expose a term that should switch on or off, an incomplete
    set of subsets, or an omitted wave fixed by the conservation equations. The run therefore accepts an
    incomplete formula before testing it across sufficiently different chambers.

    \item \textbf{The run continues to follow a disproved hint.}
    Under the false-hint condition, exact evaluations refute the prescribed
    ratio-of-polynomials form. Some runs continue fitting that type of formula;
    others reject it and stop before constructing a formula for the different chambers.
    In both cases, the prompt outweighs the numerical evidence and prevents the
    run from developing the required chamber-based formula.
\end{enumerate}

These mechanisms are clearest in the no-hint condition, where four runs find
one-chamber formulas but stop between K5 and K7, while two assemble a partial
alternating subset sum but still miss cases at K7--K8. The following case study
compares four of these runs that share the same early calculations but stop at
different points.

\subsection{Case study: four runs with no hints}
\label{sec:four-no-hint}

The four selected runs stop at two different points. Codex 5.5 and Fugu combine
the chamber terms into one formula, but each formula still misses some cases.
Claude max and Claude ultra derive correct formulas in several chambers, try to
combine them, and then report only the principal-chamber formula after that
attempt fails. Their early steps are nearly identical, so we state the common
calculation once before comparing their final results.

All four construct an independent Berends--Giele recursion evaluator, recover the
fact that the amplitude is imaginary and scales as a fixed power of frequency,
and find the one-term principal-chamber formula. Define
\begin{equation}
m=n-3,\qquad
U=\beta^2=\min(\omega_1^2,\omega_2^2),\qquad
x_j=\omega_j^2
\end{equation}
for each plus wave $j$. After removing the common factor, write
\begin{equation}
\widehat A_n
=
\frac{g^{n-3}A_n}
     {i\,2^{n-1}\omega_1\omega_2}.
\end{equation}
In the principal chamber, every run finds $\widehat A_n=U^m$. Each then produces an
explicit counterexample and concludes that the answer changes across
chambers.

\begin{figure}[t]
\centering
\includegraphics[width=0.72\linewidth]{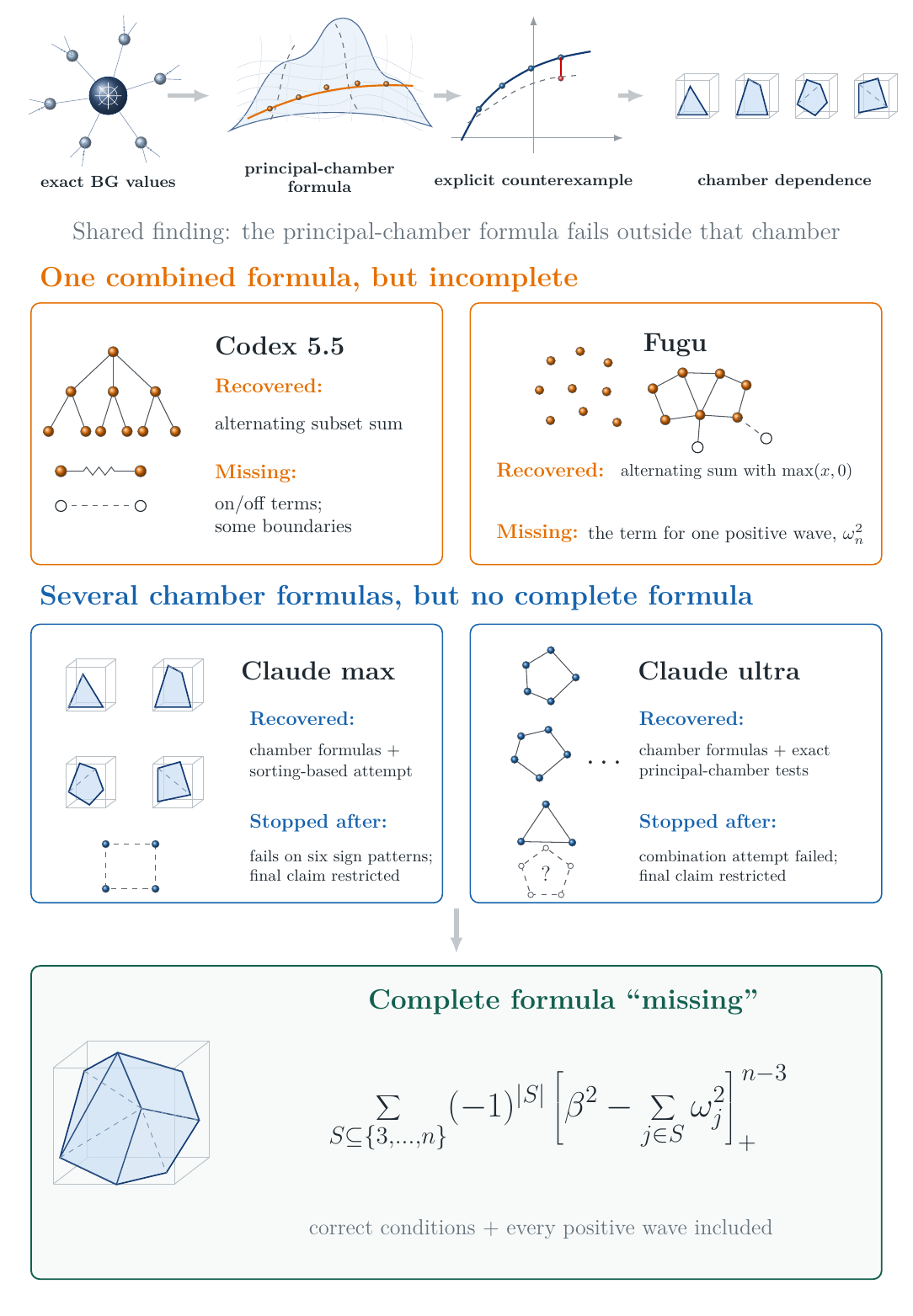}
\caption{Four representative runs without human hints all recover the
one-term formula in the principal chamber, where no chamber-boundary
correction has yet switched on, and find counterexamples outside that chamber.
Codex 5.5 and Fugu then produce short but incomplete subset sums. The Claude
runs also attempt one formula for all chambers: Claude max's sorting-based
formula misses six of 233 tests, while Claude ultra cannot combine its separate
chamber expressions. Both then report only the principal-chamber formula. The
bottom expression omits the common factor
$i\,2^{n-1}g^{3-n}\omega_1\omega_2$. Thus, these runs diagnose chamber
dependence but fail to synthesize and verify one globally valid formula.}
\label{fig:four-run-comparison}
\end{figure}

At five points, the approaches diverge. Codex 5.5
and Fugu reorganize successive pieces for two relevant positive-wavenumber squares $x_1$
and $x_2$ into
\begin{equation}
U^2,\qquad
U^2-(U-x_1)^2,\qquad
U^2-(U-x_1)^2-(U-x_2)^2+(U-x_1-x_2)^2.
\end{equation}
The last term corrects the overlap between the two single-wave subtractions.
The $\max(x,0)$ operation in the following expression automatically turns
each correction on only where it is needed:
\begin{equation}
T_m(U;\{x_j\})
=
\sum_{S}(-1)^{|S|}
\pp{U-\sum_{j\in S}x_j}^{m}.
\end{equation}
Claude max and Claude ultra derive a different formula for each sampled
chamber. For each chamber, they choose a reference point, determine the signs
of the absolute-value terms there, and simplify the recursion using those fixed
signs. Each resulting formula applies only in that chamber. They stop before
finding a single expression that gives the correct result for arbitrary
frequencies, so their final answers report only the principal-chamber formula.
Codex 5.5 and Fugu go further by recognizing the subset-sum pattern above,
although each leaves one part incomplete.
Figure~\ref{fig:four-run-comparison} summarizes these two stopping points.

\paragraph{What each formula misses.}
Codex 5.5 replaces each $\max(x,0)$ factor by an ordinary power. Its tests use
chambers in which those two expressions agree, so they do not reveal the
error. Fugu uses the correct condition for including each term but leaves out
one wave whose positive wavenumber is fixed by the conservation equations. Its
test points rarely require that missing term. Claude max proposes a formula
based on the two smallest frequency magnitudes, but the formula fails at six of 233 test
points. Claude ultra derives separate formulas in several chambers but cannot
combine them into one expression. The two Claude runs then report and test only
the principal-chamber formula. Together, the four runs show three requirements
for a complete result: use a formula built from the observed boundaries, revise
it when a test fails, and keep testing the original task of arbitrary allowed
frequencies. Appendix~\ref{app:no-hint-details}
records the per-agent formulas, test counts, counterexamples, and implementation
details.

\subsection{From single-agent failures to a coordinated multi-agent system}

The four no-hint runs complete the same early work but leave two different gaps.
Claude max and Claude ultra derive correct formulas in several chambers but do
not combine them, whereas Codex 5.5 and Fugu produce a single short formula but
do not test it broadly enough to expose an incorrect power or a missing wave.
Completing the task therefore requires both cross-chamber synthesis and tests
chosen outside the cases that suggested the formula.

In a single-agent run, the same agent usually chooses both the formula and its
test points. The assumptions
used to build the formula can therefore also shape the tests. A candidate may
appear correct because the test points repeat the same cases that suggested
the formula. Testing in deliberately different chambers is needed to reveal
the missing cases.

The PI assigns these jobs to two students. The students
try different ways to derive the answer and record their formulas,
assumptions, failed examples, and computer checks in a shared document. The PI
then writes separate code for the exact BG calculation and each proposed
formula. The PI checks that every wave with positive wavenumber is included
and chooses test points from chambers that were not used to construct the
formula. Every failed test becomes a specific task for the next round.

The required follow-up tests can be stated explicitly. For the Codex 5.5
formula, the PI should choose two squared frequencies $x_1$ and $x_2$
that satisfy
\begin{equation}
x_1,x_2<U,
\qquad
x_1+x_2>U,
\end{equation}
where $U=\min(\omega_1^2,\omega_2^2)$ was defined above. This choice tests a
case in which the terms for $x_1$ and $x_2$ are both nonzero but the term
involving $x_1+x_2$ is zero. For Fugu, the PI should swap which wave has
positive wavenumber and choose values for which the omitted wave must
contribute. For Claude max, each failed test should be used to find a missing
condition under which the formula changes. For Claude ultra, the separate
chamber formulas should be combined with factors of the form $\max(x,0)$,
which include each term only where it applies. These tests follow directly
from the observed errors and do not reveal the final formula.
Section~\ref{sec:multiagent} asks whether the PI--student team completes these
remaining steps.

\section{A coordinated multi-agent system}
\label{sec:multiagent}

The single-agent failures motivate a team process that separates formula
construction from cross-chamber testing. We first test whether this process can
rediscover the two-minus formula without hints, then apply it to the new
three-minus problem and compare four team setups.

\subsection{How the team works}
\label{sec:multiagent-workflow}

The team repeats three steps: propose a formula, compare it with the exact BG
calculation, and revise it when the two disagree. Other studies use similar
test-and-revise processes \citep{suttonbarto2018,madaan2023selfrefine,
Shinn2023Reflexion,Gou2024CRITIC,Wu2024ProCo}. After each round, the agents add
their formulas, failed examples, computer checks, and next tasks to one shared
record. Panel (c) of Figure~\ref{fig:paper-overview} shows these three steps.

The team has one PI and two students. All three receive the same
question, exact BG code, and list of tests that the final answer must pass.
They can all read the shared record, but each writes and runs code in a separate
workspace. The PI assigns the tasks, combines the results, and prepares the
final answer. This arrangement lets one agent propose an answer while another
checks it, as in earlier teams of agents
\citep{Li2023CAMEL,Wu2023AutoGen,Qian2023ChatDev,Hong2023MetaGPT,
Du2023MultiagentDebate,Guo2024MultiAgentSurvey,Schmidgall2025AgentLaboratory,
Gottweis2025CoScientist,VillaescusaNavarro2025Denario,Ghareeb2026Robin}.

The written instructions state what each agent is responsible for, which files
to use, which tests are required, and when the PI may stop. They do not tell
the PI which formula to try or which extra values to test. The PI chooses
those from the results of the current round. Appendix~\ref{app:pi-student-prompts}
reproduces the complete instructions used in both two-minus runs.

At the start of a round, the PI identifies the most important missing piece
and gives the students different questions. In our runs, one student
usually calculated exact examples while the other tried to derive and explain
a formula. Both placed their results in the shared record. The PI then wrote
separate code for the proposed formula and compared it with the exact BG
calculation. If a test failed, that failure became a task in the next round. If
all required tests passed, the PI ended the run.

For the two-minus experiments, the PI had to test $n=4,5,6,7$ at several
frequency choices for each $n$, including choices with very different
magnitudes. The formula and BG calculation had to agree to a relative error of
at most $10^{-10}$. This setup directly addresses the failures in
Section~\ref{sec:four-no-hint}: the students search for the formula, and
the PI checks failed examples, missing cases, and whether the answer is
complete.

\subsection{Rediscovering the complete two-minus formula}
\label{sec:multiagent-two-minus}

We applied this team process twice to the two-minus task, once with Claude Opus
4.8 agents and once with Codex 5.5 agents. These experiments parallel the six
no-hint single-agent runs in Section~\ref{sec:single-prompt-results}: both
settings provide the same problem and BG recursion evaluator without hints,
literature access, or an answer key. None of the six single agents finds the
all-chamber formula. Both teams, by contrast, recover it automatically within
two or three rounds.

The Opus team recovered Equation~\eqref{eq:key} in two rounds. In round 1,
one student examined exact data across frequency chambers and assembled the
alternating sum over subsets. The other student derived the
one-term principal-chamber formula and explained how it changes when every frequency is
multiplied by the same number. Their results
agreed where they applied to the same frequencies. In round 2, the PI implemented the
candidate independently and obtained 142 exact matches, including 95 points
outside the principal chamber, then stopped the run.

The Codex 5.5 team reached the same formula in three rounds through a different
path. Its first round ruled out simple symmetric candidate polynomials. In
round 2, one student showed that the special $n=4$ limit gives the same answer
regardless of how it is approached, while the other reconstructed the
alternating subset sum chamber by chamber. In round 3, the
PI found exact agreement on 12 tests away from the limit and checked 12 approaches to
the $n=4$ limit from points that do not yet satisfy the conservation equations;
the largest discrepancy was
$2\times10^{-12}$, below the stated $10^{-10}$ threshold.

\begin{figure}[t]
\centering
\includegraphics[width=\linewidth]{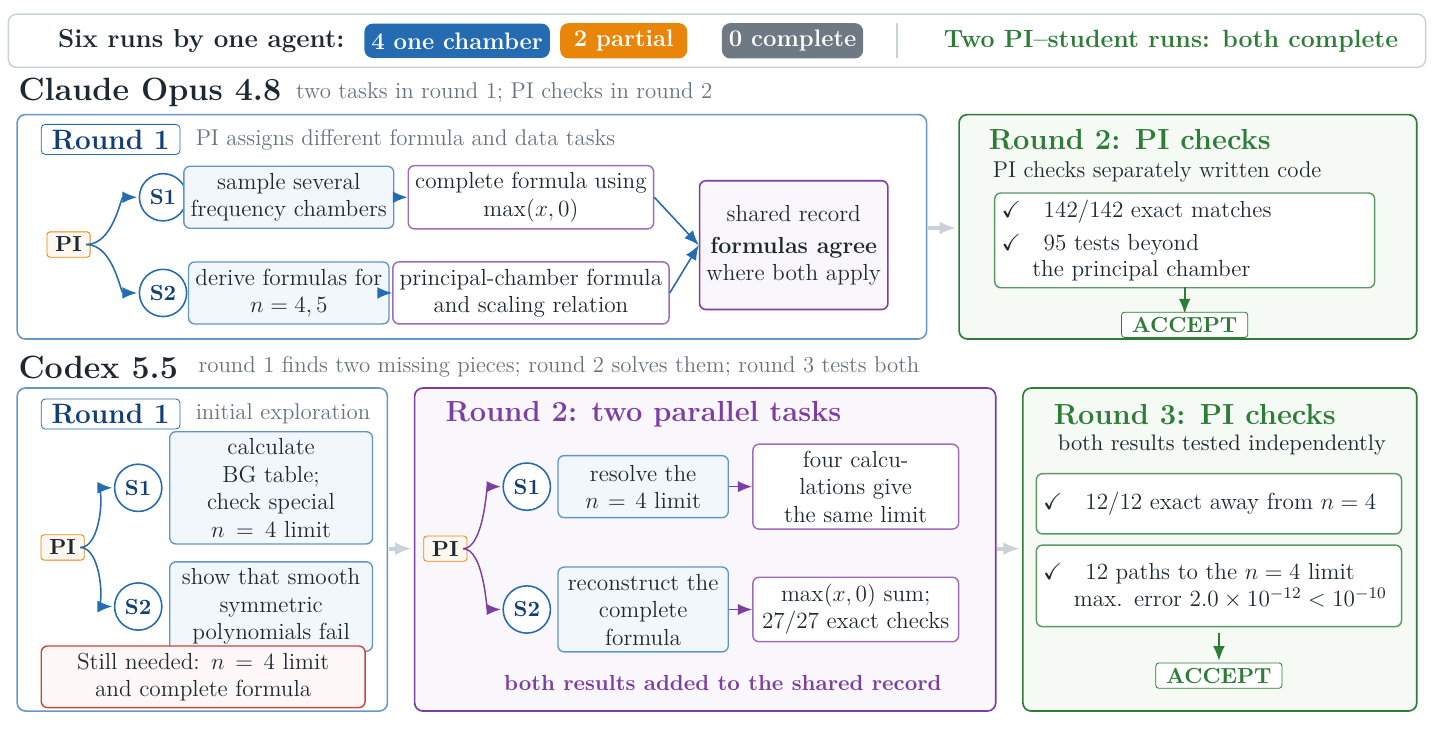}
\caption{Workflow of our PI+two student (S1, S2) architectures with different base models. The two students construct formulas, while a PI assigns work and
checks the result independently. The top row shows the six no-hint
single-agent results for comparison, and each box records one team round. The
Opus team proposes the complete formula
in round 1 and verifies it in round 2; the Codex 5.5 team uses its first failure
to divide round 2 between the special $n=4$ limit and the complete formula,
then the PI independently tests both results and accepts them in round 3. Thus
both coordinated teams succeed where none
of the six no-hint single agents does, although the runs are not cost-matched
and are too few to measure success rates. A written record, two simultaneous
searches, and separate final checks help the teams complete the formula.}
\label{fig:multiagent-comparison}
\end{figure}

The two histories in Figure~\ref{fig:multiagent-comparison} close different
gaps left by the single-agent runs. In the Opus run, the PI compares the formula
inferred from data with an independent derivation in the principal chamber. In
the Codex 5.5 run, the students separately resolve the complete formula and the
exceptional $n=4$ boundary. In both cases, the PI then tests the combined claim
outside the chambers used to construct it.

\subsection{New formula: the six-point three-minus amplitude}
\label{sec:multiagent-three-minus}

The two-minus experiments show that the team can rediscover a known target
without seeing its answer. To test whether the workflow can also produce a new
result, we turn to the
\emph{three-minus sector}, in which three wavenumbers point in the negative
direction,
$\sigma=(-1,-1,-1,+1,\ldots,+1)$. At five points, reversing every
wavenumber maps this case back to the two-minus sector. Six points is the first
number of waves for which three-minus scattering is genuinely distinct and
contains a richer set of chamber boundaries. To our knowledge, no closed
form for $A_6$ in this sector was previously known. The expression below is
therefore the main new scientific result of the team, with its tests and limits
stated below.

Unlike the controlled rediscovery runs, this experiment allowed literature
access, used more compute, and continued for all scheduled rounds. We gave the
agents two known facts: the vanishing one-minus sector and Eq.~\eqref{eq:key}.
We also noted that reversing every wavenumber sign maps the five-point
three-minus problem to the known two-minus problem.

\begin{figure}[H]
\centering
\includegraphics[width=0.98\linewidth]{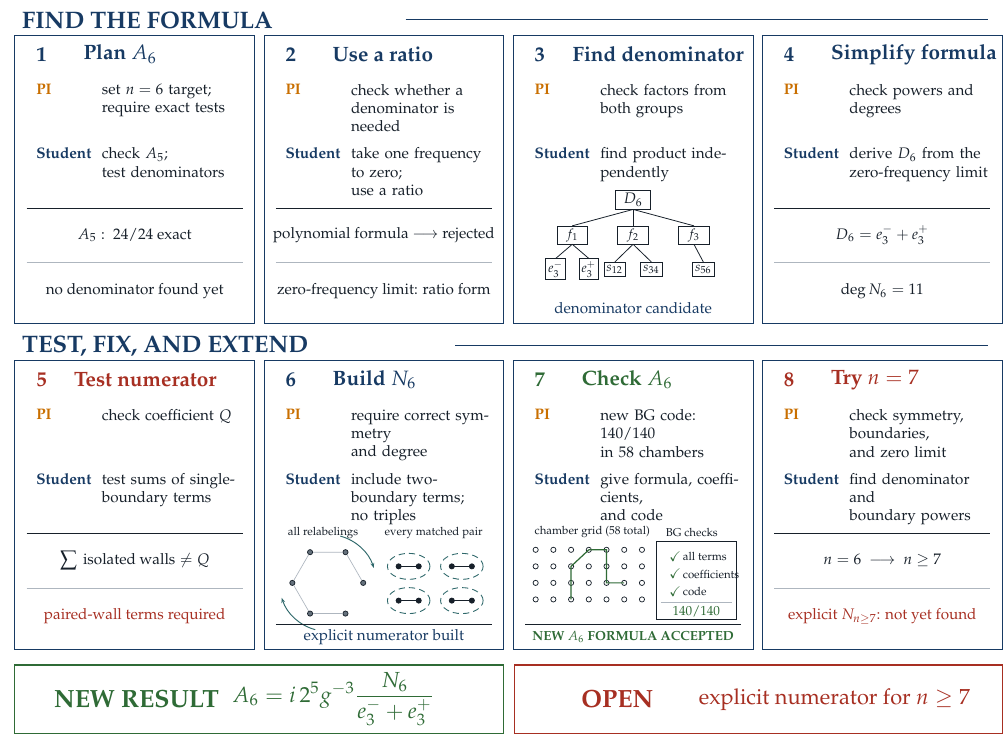}
\caption{After rediscovering the known hydrotope formula for two negative
wavenumbers, the agents tackle the harder sector with three negative
wavenumbers and propose a new analytic expression for the six-point amplitude
$A_6$ intended to cover every chamber (see Eq.~\eqref{eq:A6-three-minus}).
% The six-point three-minus amplitude is the first three-minus case that differs from the known two-minus problem.
In this eight-round run, rounds 1--4
determine the formula's ratio structure: a common frequency-dependent
denominator and a polynomial numerator carrying the remaining frequency
dependence. A failed test in round 5 reveals that the numerator needs
corrections involving pairs of chamber boundaries. Rounds
6--7 add those terms, and independently written BG recursion code finds exact
agreement on 140 tests in 58 chambers. Round 8 studies $n=7$, whose explicit
numerator remains unknown. The failed cross-chamber test drives the correction
that completes the supported $A_6$ expression, but the finite tests do not
provide an analytic proof and the numerator for $n\geq7$ remains open.}
\label{fig:three-minus-a6}
\end{figure}

The discovery proceeded through a sequence of increasingly global candidates.
The group first determined $A_5$, then
showed that $A_6$ could not be one polynomial. Two students independently found
the same denominator for a ratio of polynomials. A later candidate containing
corrections for one chamber boundary at a time failed: the differences from the BG
recursion showed that
some corrections must appear together at pairs of chamber boundaries.
The team therefore enlarged the numerator to an expression that uses different
polynomials in different chambers and includes corrections involving two
boundaries at once. Figure~\ref{fig:three-minus-a6} summarizes this progression.

Let
\begin{equation}
M=\{1,2,3\},
\qquad
P=\{4,5,6\},
\qquad
a_i=\omega_i^2\ (i\in M),
\qquad
b_j=\omega_j^2\ (j\in P),
\end{equation}
and define
\begin{equation}
G_0=S_3(M)\times S_3(P),
\qquad
G=G_0\rtimes\langle(1\,4)(2\,5)(3\,6)\rangle .
\end{equation}
For $H\subseteq G$, we use the unnormalized group sum
\begin{equation}
 S_H[F](\omega)
\equiv
\sum_{\pi\in H}F(\pi\cdot\omega),
\qquad
(\pi\cdot\omega)_r=\omega_{\pi^{-1}(r)}.
\end{equation}
The six-point amplitude is
\begin{equation}
A_6
=i\,2^5g^{-3}\frac{N_6}{e_3^-+e_3^+},
\qquad
e_3^-=\omega_1\omega_2\omega_3,
\qquad
e_3^+=\omega_4\omega_5\omega_6,
\label{eq:A6-three-minus}
\end{equation}
with
\begin{equation}
\begin{aligned}
N_6={}&B+ S_G\!\Bigl[
  (b_4-a_1)_+P_{14}
  +(b_4-a_1)_+(b_5-a_2)_+R_{14,25}
  +\frac18(a_1-b_4-b_5)_+^3Q_{145}
  \Bigr].
\end{aligned}
\label{eq:N6-three-minus}
\end{equation}
Here $[x]_+=\max(x,0)$.  The sums \( S_H\) are sums over group
elements, not averages and not sums over distinct walls.  Consequently the
single-wall seed occurs with its four-element spectator stabilizer, while the
ordered matching seed has no additional stabilizer factor.  On the conservation
locus, class exchange maps each three-leg wall to its complementary description,
so the full group counts each physical $Q$ term eight times; this accounts for
the factor $1/8$.  Appendix~\ref{app:a6-three-minus} gives
the four seed polynomials $B$, $P_{14}$, $R_{14,25}$, and $Q_{145}$ and expands
the group sums into ordinary indexed sums.

As a final independent check, the PI generated every relabeled term and tested
the formula using separately written BG recursion code with exact fraction
arithmetic. This run resolves only the three-minus case for $A_6$; constructing
a formula for $n\geq7$ remains future work.

\subsection{Comparing four team setups}
\label{sec:architecture-comparison}

The teams perform better in the runs above, but those results do not show which
part of the team process matters most. We therefore compare four
team structures on one target: does the run end with an explicit formula for
$A_6$ that resolves its chamber dependence and passes tests in kinematic
chambers not used to construct it?

\begin{enumerate}
    \item \textbf{Single agent.} Only one agent and a single session.

    \item \textbf{PI $+$ 1 student.} The team process of
    Section~\ref{sec:multiagent-workflow} with only one student.

    \item \textbf{PI $+$ 2 students.} The team process of
    Section~\ref{sec:multiagent-workflow} with two students.

    \item \textbf{PI $+$ 2 students $+$ verifier.} The
    PI--student team with a verifier assigned to check the result
    independently. The prompt tells the PI not to perform these checks
    because they belong to the verifier.
\end{enumerate}

We ran each setup once, so the results describe these four runs rather than
their general success probabilities. The two PI$+$two-student setups used the
same scientific prompt, round schedule, and PI/student models, but each PI chose
tasks from the evidence in its own run. The comparison therefore includes both
a different number of agents and different task choices.

The PI in the last setup departed from its prompt: it was told not to check
the formula, but nevertheless ran 5,733 tests while developing and combining
candidates. The verifier ran more than 6,000 tests using
independently written code and returned the failures to the team. This run
therefore contains two sources of checking, so it does not show what would
happen if only the verifier performed that work.

\begin{figure}[t]
\centering
\includegraphics[width=0.98\linewidth]{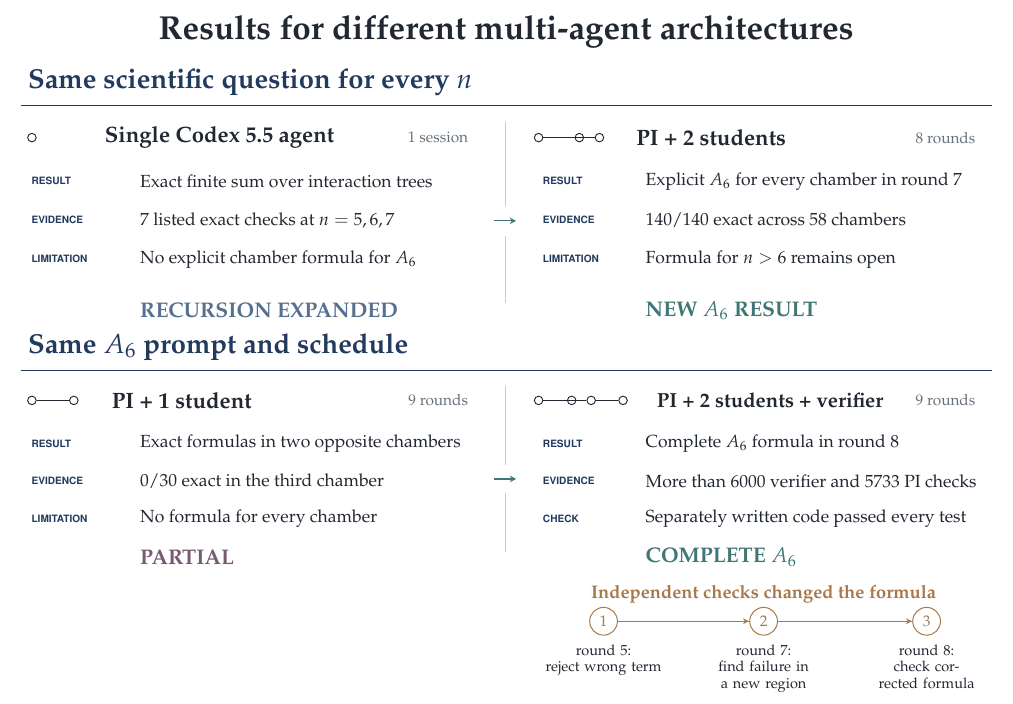}
\caption{AI agents often find formulas that work in one frequency chamber but
fail to combine them into a global expression. We compare four architectures
on the harder discovery problem introduced in
Figure~\ref{fig:three-minus-a6}: deriving an explicit formula for the new
six-point amplitude $A_6$ in the three-negative-wavenumber sector and testing
it in unseen chambers. The single-agent run returns a tree expansion that hides
the chamber changes, while the PI$+$one-student team finds separate chamber
formulas that fail in a new chamber. Both PI$+$two-student setups, one with an
additional verifier, complete and independently check the global expression.
Because each architecture was run once and the runs are not cost-matched, these
outcomes identify failure modes rather than success rates. In these runs, only
teams with two complementary student roles carry both formula construction and
cross-chamber testing through to a complete $A_6$ formula.}
\label{fig:a6-architecture-comparison}
\end{figure}

The two unsuccessful runs in
Figure~\ref{fig:a6-architecture-comparison} stop for different reasons. The
single agent returned a valid nonrecursive expression: it expanded the BG
recursion into a finite sum over tree-shaped interaction histories and matched
all seven points it checked. It did not, however, produce the requested
explicit formula for $A_6$. Absolute values of intermediate momentum sums
still hide where the expression changes from one polynomial to another. None
of the seven test points lies near a chamber boundary, making the changes hard to
detect. This run illustrates a concrete risk of having the same agent choose
both the construction data and the test data: its tests can repeat the same
limited cases that suggested its expression.

The one-student team stopped at a different point. It addressed the chamber
structure directly and derived independently checked formulas in two opposite
frequency chambers. Neither formula extended to a third chamber: both failed on
all 30 new test points, and the discrepancy remained unresolved when the run
ended. The team therefore failed to combine its two chamber formulas into one
complete formula. This occurred despite a PI, a written record, and a
separate implementation of the calculation. With only one student, each
round forced a choice between extending the equation and generating evidence
from other chambers, so the team could not advance both parts of the task in
parallel.

The two-student teams could develop a formula and test it at the same
time. In one run, one student explored different frequency chambers
while the other searched for a common pattern. In the run with a
verifier, the verifier rejected an incorrect boundary correction in round 5 and found a
failure away from the chamber boundaries in round 7; the
team then revised the formula, which passed the independent check in round 8.
In both cases, cross-chamber evidence directly shaped the formula developed in
later rounds.

\section{Discussion}
\label{sec:discussion}

The final answer alone does not show how much of the problem an agent solved.
Fifteen of the 18 runs found the principal-chamber formula, 15 recognized that
different chambers require different polynomials, and eight found every
boundary where the formula changes. Most runs stopped while combining those
chamber formulas or while testing the combined result. Evaluations should
therefore record the intermediate formulas, counterexamples, claimed range of
validity, and test points, rather than placing every incomplete answer in one
category.

One source of divergence is that the prompts changed both the formulas
considered and the examples tested. The
true hint told agents to expect a different polynomial in each chamber, and
four agents then found the complete formula. The false hint proposed one ratio
of polynomials and recommended frequencies that rarely expose chamber changes.
Some agents rejected that suggestion after exact counterexamples; others kept
following it. Human suggestions can be tentative, incomplete, or outdated, so
a useful scientific agent must revise both its formula and the assumptions in
its prompt when exact calculations disagree.

Coordination addresses a different source of failure: the same agent no longer
has to construct the formula and choose all of its tests. In the
PI$+$student teams, two students pursued analytic and numerical questions in
parallel, and their written record preserved formulas and failed tests. The
PI, or the verifier in one setup, implemented the calculation
again and tested the combined formula. This caught errors that agreement among
the proposing agents had missed. Both PI--student teams received
no formula hint and found the complete hydrotope formula, while none of the six
comparable single-agent runs did. The teams used more model calls, so these runs
do not show that adding agents alone caused the improvement.

Beyond supporting parallel work, the shared record let each student focus on
the evidence needed for its task rather than reread the entire investigation.
Such task-focused context can help when long histories make relevant
information harder to use
\citep{Liu2023LostMiddle,Hsieh2024RULER}. This benefit does not imply that
larger teams are always better: a very large team could repeat work and make
communication harder. The practical goal is the smallest team that can pursue
independent tasks and check the final result.

The appropriate method also depends on the scientific goal. Standard machine learning is
appropriate when accurate prediction on similar data is the goal. Symbolic
regression is appropriate when the variables and allowed operations are known
and an explicit equation is desired. An agent is useful when the work also
requires choosing new variables, finding where a formula changes, and selecting
new calculations that separate competing explanations. In this problem, adding
a language model to symbolic regression still left the search driven mainly by
average error. The PI--student team instead used individual counterexamples to
change the proposed formula and combine the chamber results.

The advantage of interactive revision also relies on having an exact way to
check a proposed formula. BG
recursion provides that check here, and accurate simulations can play the same
role in related problems \citep{guevara2026singleminus,
Islam2026InterpretableSurrogates,Islam2026UnifiedRemnant,
Schwartz2026CParameter}. Exact checks quickly reject a wrong formula, but they
do not supply the correct formula or prove that finitely many successful tests
cover every case. The new six-point three-minus expression shows that the same
propose--test--revise process can produce a previously unknown result. Its many
exact tests provide strong evidence, while an analytic proof and an extension
to arbitrary $n$ remain open.

The main limitations of the present comparisons point to two follow-up
experiments. First, future tests
should give single agents and teams of agents the same total computational
resources and repeat each setup enough times to estimate reliability
\citep{Guo2024MultiAgentSurvey,Chen2024ScienceAgentBench,
Lu2025AgentRewardBench}. Second, the shared written record should connect each
proposed formula with its supporting examples, counterexamples, code, and open
questions. Agents can then retrieve only the material needed for the current
task and avoid repeating earlier work.

\section{Conclusion}
\label{sec:conclusion}

We reconstructed the original human--agent discovery of the hydrotope,
repeated the task in 18 single-agent runs, and compared those runs with machine
learning and symbolic regression. The difficult step was combining formulas
from different frequency regions and testing the result in new regions. A
PI and two students completed this step in two no-hint runs. The same
team also found a new expression for the six-point three-minus amplitude. When
an exact calculation or accurate simulation is available, scientific agents
should use counterexamples to revise proposed formulas and should have a
separate check before accepting the final result.

\section*{Acknowledgments}

We thank Sid Mishra Sharma and Ioana Ciucă for useful discussions. The authors acknowledge the Texas Advanced Computing Center (TACC) at The University of Texas at Austin for providing computational resources that have contributed to the research results reported within this paper. URL: http://www.tacc.utexas.edu. This work was supported by the National Science Foundation under Cooperative
Agreement 2421782 and the Simons Foundation award MPS-AI-00010515
(NSF-Simons AI Institute for Cosmic Origins - CosmicAI, \url{https://www.cosmicai.org/}. MZ acknowledges support
from the National Science Foundation NSF-BSF 2207583
and NSF 2209991, the Nelson Center for Collaborative
Research and the Simons Foundation through the Black
Holes and Strong Gravity program through Award No.
SFI-MPS-BH-00012593-10

\appendix

\section{Prompts used to repeat the discovery task}
\label{app:prompt-packages}

Table~\ref{tab:prompt-packages} summarizes the three prompts. This appendix
records their equation and sampling guidance in full because the experiment
varies the complete prompt rather than
the formula hint alone.

\conditionbox{\case{case\_1}: false hint}{falsehintborder}{falsehintbg}{
\textbf{Prompt hint:} \emph{``The amplitude $A_n$ is a rational function (a
ratio of polynomials) of the
frequencies $\{\omega_i\}$ -- a single global, analytic expression valid
throughout the entire two-minus sector.''} The prompt recommends one formula
$N(\omega)/D(\omega)$ for the entire two-minus sector, with denominator factors suggested by
intermediate wave interactions. It states that there is no chamber
decomposition, no absolute values,
and no $\min/\max$, and says the answer is emphatically not a plain polynomial.
It also recommends sampling generic comparable frequencies and deliberately
avoiding widely separated frequencies or points close to chamber boundaries.
}

This package tests whether an agent can reject the supplied single
ratio-of-polynomials description when it conflicts with numerical evidence,
despite sampling advice that steers the search away from informative, widely
separated frequency scales. A successful run must abandon the proposed
ratio-of-polynomials form and discover the correct explicit formula.

\conditionbox{\case{case\_2}: true hint}{hintborder}{hintbg}{
\textbf{Prompt hint:} \emph{``The amplitude in the two-minus sector is a
piecewise homogeneous polynomial in the frequencies $\{\omega_i\}$.''} Here,
``homogeneous'' means that rescaling all frequencies by the same factor rescales
the answer by a fixed power. The prompt further states that the answer contains
neither ratios nor functions such as exponentials and logarithms, and that a
different polynomial applies in each frequency chamber.
}

This package states correctly that a different polynomial applies in each chamber, but does not give
the answer itself. It tests whether the agent can use that description together
with numerical data to identify the chamber structure and reconstruct a single
formula valid in every chamber.

\conditionbox{\case{case\_3}: no hint}{nohintborder}{nohintbg}{
\textbf{Prompt hint:} none. The prompt contains the physical setup,
the BG recursion code description, the two-minus sector, and the task:
\emph{``Find a closed-form analytic formula for $A_n$ in the two-minus sector,
valid for all $n\geq4$ and for arbitrary kinematics in this sector.''}
It asks for numerical evidence at $n=4,5,6,7$ and explicitly includes
widely separated frequency scales, e.g. one frequency much larger or much
smaller than the others.
}

This no-hint package tests whether the agent can discover the answer without
a proposed formula class. Its request for tests at separated scales
distinguishes it from the sampling advice in the other conditions.

\section{Details of the four no-hint runs}
\label{app:no-hint-details}

Section~\ref{sec:four-no-hint} compares four runs that share the same early
steps but stop at two different stages. Here we give the per-agent
formulas, test counts, counterexamples, and implementation details underlying
that comparison.

\subsection{Two partial sums and what each misses}

Codex 5.5 finds the alternating subset sum by examining exact chamber
polynomials at $n=5$ and $n=6$. Its final expression keeps only the $r$
smallest individual squares below $U$ and uses an ordinary power where a
$\max(x,0)$ is needed. It works in the nested chambers used to derive and
test it, but fails when two individual squares are below $U$ and their sum is
above $U$. For example, let $U=16$ and $x_1=x_2=9$. Each $x_j$ is below $U$,
but $x_1+x_2=18$ is above it, so the overlap term must be zero. None of the 20
test points includes this case. The final formula therefore has the right
alternating sum but the wrong condition for turning one term on or off.

Fugu keeps the $\max(x,0)$ terms and derives the alternating sum directly from
the sign changes of the factors
$|k_S|=|\sum_{i\in S}\sigma_i\omega_i^2|$, the magnitudes of intermediate
wavenumber sums. Its boxed formula passes 500 random tests at each
$n=4,\ldots,8$ using the standard sampling routine. The remaining defect is
the index set: Fugu sums over $R=\{3,\ldots,n-1\}$ and omits the term
$\omega_n^2$ for one positive-wavenumber wave. The evaluator fixes this term
through the conservation equations. A
method used to generate the sampled chambers hides the omission, but changes
or becomes undefined in other chambers. Thus Codex 5.5 uses the wrong condition
for including a term, whereas Fugu uses the right condition but leaves out one
wave with positive wavenumber.

\subsection{Two runs that finish with the principal-chamber formula}

Claude max and Claude ultra both freeze the absolute-value signs at selected
reference points and obtain several correct five-point chamber formulas.
These expressions are organized in their chosen free-frequency coordinates by
wave orderings and sign patterns; neither run reorganizes them into the complete
subset-wall sequence displayed in Section~\ref{sec:four-no-hint}. Both identify the symmetric principal-chamber
formula
\begin{equation}
A_n^{\rm principal}
=
i\,\frac{2^{n-1}}{g^{n-3}}\,
\omega_1\omega_2
\left[\min(\omega_1^2,\omega_2^2)\right]^{n-3}.
\end{equation}
Each run tries to combine its chamber formulas, but neither tries the
alternating subset sum. After that attempt fails, each reports the
principal-chamber formula even though the prompt asks for arbitrary allowed
frequencies.

Claude max attempts to choose among the chamber formulas using the two waves
with the smallest frequency magnitudes. The formula passes 227 of 233 random tests and fails
on six configurations with previously untested frequency-sign patterns.
Keeping the absolute values unresolved in symbolic simplification also
exhausts the available memory. The run interprets these outcomes as evidence
that the remaining chambers require separate cases, calls the principal
chamber the physically relevant region, and returns its one-term formula after more
than 110 tests inside that chamber. The run therefore changes the question
after its attempt to combine the chamber formulas fails.

Claude ultra supplies an especially strong check of the BG calculation. It
builds independent Python code using fractions rather
than floating-point approximations, finds and repairs a missing factor in its
first implementation, and then matches the Mathematica evaluator through
$n=8$, including a controlled $n=4$ limit. It maps several non-principal
five-point formulas and attempts to combine them into one absolute-value
expression; the proposed combination is discontinuous, and the symbolic
simplification stalls. It then describes the principal
chamber exactly by $|\omega_2|=\min_i|\omega_i|$ and verifies the equivalence
between that condition and the one-term formula on 79 random points. Those tests
show exactly where the formula works, but the run does not use the failures
outside that chamber to correct the formula. It therefore restricts the answer
after an unsuccessful combination attempt instead
of constructing the alternating subset sum.

\section{Details of the symbolic-regression comparisons}
\label{app:sr-diagnostics}

The conventional SR methods use three random initializations (seeds) for
PySR and gplearn and ten for Operon. At $n=5$, PySR uses 35 iterations with 12
populations of 24 candidate equations, gplearn
uses a population of 2,000 for 60 generations, and Operon uses a population of
5,000 with at most two million evaluations. At $n=7$, we enlarge the corresponding PySR
and gplearn searches to 150 iterations with 20 populations of 40
and a population of 3,000 for 100 generations, respectively; the Operon
evaluation cap remains unchanged. In versions that are told which operations
they may use, PySR and LaSR receive arithmetic, squaring, minimum, and
$\max(x,0)$, while gplearn and Operon receive arithmetic, squaring, minimum,
and maximum. Thus PySR and LaSR may use $\max(x,0)$, but no
method receives precomputed inputs such as
$[\beta^2-\sum_{j\in S}\omega_j^2]_+$. Each method must discover the relevant
subset sums, the quantities placed inside each operation, and the combination of terms from the raw
frequencies.

The additional $n=5$ comparison uses only raw frequencies. Bingo receives
arithmetic, absolute value, and squaring, with three runs capped at 500,000
evaluations \citep{Randall2022Bingo}. PS-Tree uses arithmetic, minimum,
maximum, analytical quotient, sine, cosine, and up to eight learned
decision-tree regions, fitting a separate equation within each region
\citep{Zhang2022PSTree}. A separate region-first method selects a regression
tree by three-fold cross-validation on the
training rows and then runs Operon independently in each leaf. None receives
the target formula, analytic chamber labels, or held-out feedback.

LaSR and LLM-SR use two random initializations for each value of $n$ with
GPT-5.6-Terra. LaSR uses 40 iterations, 20 populations of 50, and maximum
expression size 70. LLM-SR generates 100 candidates per seed across ten
separate subpopulations. Both select their reported equations using training loss before one
held-out evaluation. The standard ML methods use raw frequencies and fixed
settings: ordinary least squares; an 800-tree random forest; and a neural
network using the rectified-linear function $\max(x,0)$ (ReLU), with layer widths
$(256,256,128)$, early stopping, and at most 2,000
iterations. The KAN baseline uses canonical \texttt{pykan}~0.2.8 with raw
frequencies, training-only standardization, and cubic splines. Validation on
150 of the 750 training rows selects width 8 and grid size 5; we then refit all
750 rows for three seeds, using 50 LBFGS steps at $n=5$ and 100 at $n=7$.
Each ML method uses three seeds.

\subsection{How formulas were selected and limits of the fixed split}

After the runs finished, we chose the displayed PySR equations from the set of
formulas that trade accuracy against simplicity, using root mean squared error
(RMSE) on the held-out data. They are therefore best-case choices; the
held-out data were used for selection rather than consulted only once. Both
values of $n$ use one fixed train--test split. At
$n=7$, rare large target values make the training sample larger in scale than
the held-out sample. Variation among random seeds measures search randomness
for this split, not variation across newly drawn datasets. Run counts and
compute limits also differ across methods. The comparison therefore does not
use equal compute and cannot show that any one design choice caused a result.

\subsection{Seven-point test}
\label{app:sr-n7}

\begin{figure}[H]
\centering
\includegraphics[width=0.98\linewidth]{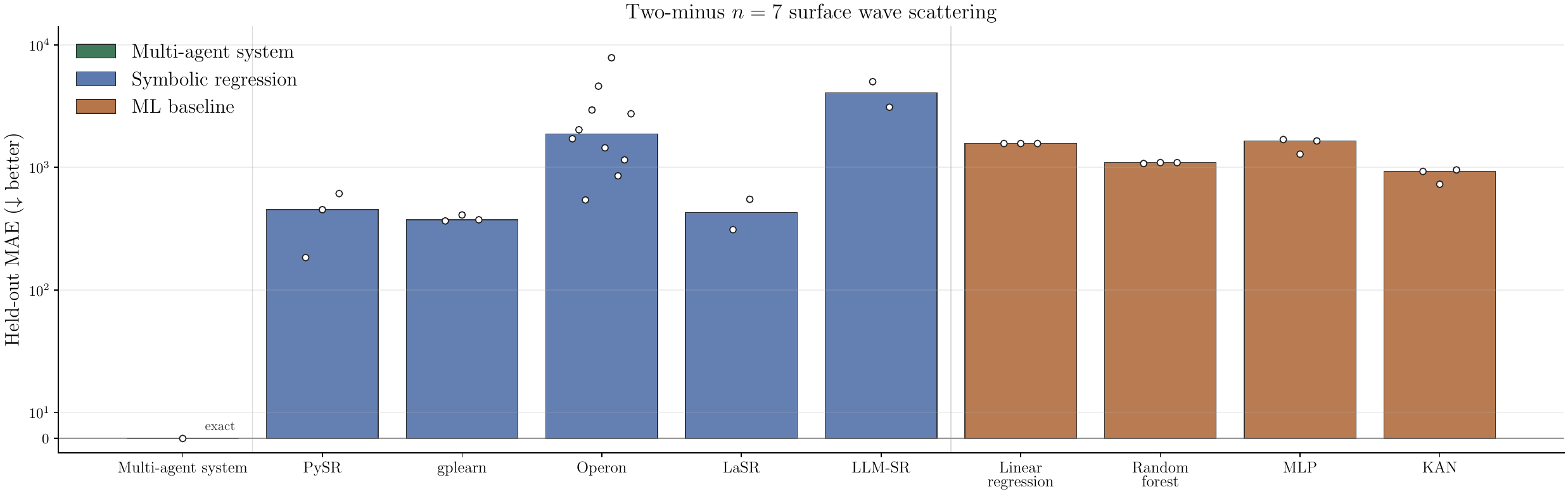}
\caption{This figure repeats the comparison in
Figure~\ref{fig:symbolic-regression-n5} for the $n=7$ case. The formula
increases from eight $\max(x,0)$ terms at $n=5$ to
32 such terms. Bars show median
held-out MAE from raw frequencies and open markers show runs; gplearn has the
lowest fitted median, 373.49, but slightly negative held-out $R^2$ and does not
recover the target equation. The research team's exact formula remains at
zero, while every fitted method misses terms that switch on in some chambers.
A few unusually large values in the fixed split also make absolute errors
larger than at $n=5$. Increasing
the number of terms that can switch on or off makes exact recovery more difficult
under the tested budgets.}
\label{fig:symbolic-regression-n7}
\end{figure}

Figure~\ref{fig:symbolic-regression-n7} repeats the same comparison at
$n=7$, where Eq.~\eqref{eq:key} contains 32 fourth-power
$\max(x,0)$ terms rather than the eight squared terms at $n=5$. gplearn has
the lowest median MAE among the fitted methods, but its held-out $R^2$ is
slightly negative and its expression is not the target formula. Most sampled
points constrain only the polynomial used in their own frequency chamber,
so reducing average error does not require recovering every chamber boundary or
the alternating sum that combines the $\max(x,0)$ terms. The larger
error scale also reflects the fixed split: the $n=7$ target occasionally takes
values much larger than its typical value,
with training mean and standard deviation approximately $990$ and $9258$, but
held-out mean and standard deviation approximately $364$ and $2729$.

\begin{figure}[H]
\centering
\includegraphics[width=0.98\linewidth]{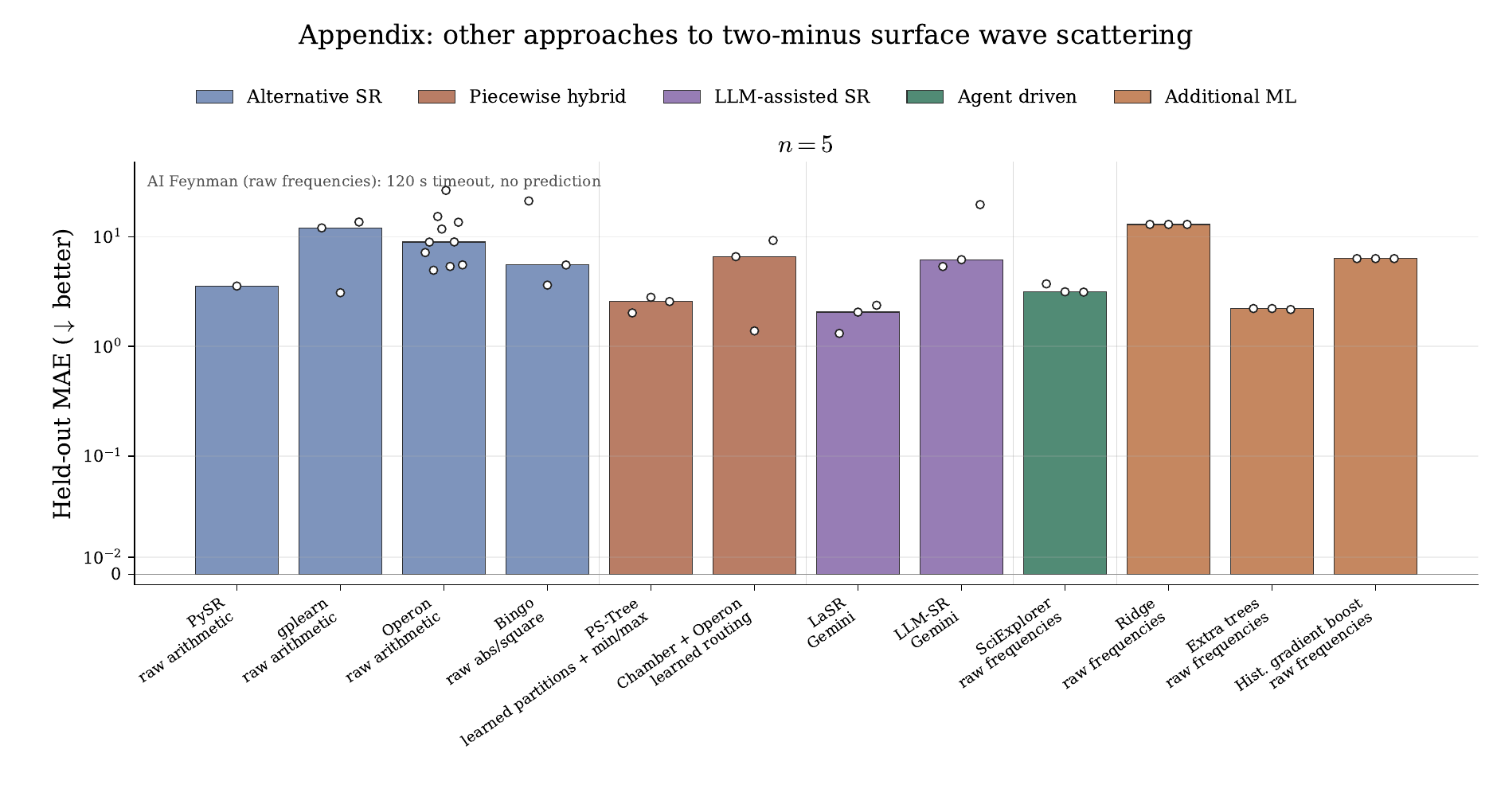}
\caption{Additional methods test whether searches that allow a wider range of
equations, including separate equations for different regions, recover the five-point formula. Bars show median held-out
MAE and open markers show individual runs for alternative symbolic-regression
systems, combinations of region-finding and equation search, archived LLM-assisted
variants, SciExplorer, and additional ML methods. PS-Tree improves approximate prediction
by learning seven regions with separate symbolic expressions, but it does not
recover the compact hydrotope formula; the region-first Operon hybrid likewise
misses the true chamber boundaries. SciExplorer misses the $\max(x,0)$ terms. Allowing separate formulas in different regions can
improve predictive accuracy, but none of these searches recovers the complete
formula found by our research team.}
\label{fig:sr-appendix}
\end{figure}

Figure~\ref{fig:sr-appendix} records additional checks omitted from the main text Figure~\ref{fig:symbolic-regression-n5}. All displayed methods use raw frequencies. The five-point figure
adds Bingo, PS-Tree, a method that first divides the data into regions and then runs Operon, additional ML methods,
archived LLM-assisted variants, and SciExplorer, an agent that searches a
fixed dataset. PS-Tree has median held-out MAE 2.57, compared
with 5.53 for Bingo and 6.59 for the region-first Operon method, but none
recovers the short formula. PS-Tree returns seven regions containing large
equations found by genetic programming, while the region-first method divides the
data mainly by frequency magnitude rather than locating the slanted chamber
boundaries. SciExplorer interacts only with training data
and training scores; none of its submitted expressions contains the
$\max(x,0)$ terms needed to switch terms across frequency chambers.

\section{Explicit coefficients for the six-point three-minus amplitude}
\label{app:a6-three-minus}

This appendix gives a compact specification of the
six-point amplitude
\begin{equation}
   A_6(1^-,2^-,3^-,4^+,5^+,6^+)
  =\frac{2^5 i}{g^3}\,\frac{N_6}{\eM + \eP},
  \label{eq:A6-master}
\end{equation}
where
\begin{equation}
  \eM=\omega_1\omega_2\omega_3,
  \qquad
  \eP=\omega_4\omega_5\omega_6,
  \qquad
  [x]_+\equiv\max(x,0).
\end{equation}
We use
\begin{equation}
  M=\{1,2,3\},
  \qquad
  P=\{4,5,6\},
  \qquad
  a_i=\omega_i^2\ (i\in M),
  \qquad
  b_a=\omega_a^2\ (a\in P).
\end{equation}

Let
\begin{equation}
  G_0=S_3(M)\times S_3(P),
  \qquad
  \tau=(1\,4)(2\,5)(3\,6),
  \qquad
  G=G_0\rtimes\langle\tau\rangle .
\end{equation}
Thus $|G_0|=36$ and $|G|=72$.  For any subgroup $H\subseteq G$, define the
\emph{unnormalized} group sum
\begin{equation}
   S_H[F](\omega)
  \equiv
  \sum_{\pi\in H}F(\pi\cdot\omega),
  \qquad
  (\pi\cdot\omega)_r\equiv\omega_{\pi^{-1}(r)}.
  \label{eq:group-sum}
\end{equation}
There is no factor $1/|H|$ in this definition. The complete numerator is most compactly and unambiguously written as the
orbit sum
\begin{equation}
\begin{aligned}
N_6={}&B+ S_G\!\Bigl[
  (b_4-a_1)_+P_{14}
  +(b_4-a_1)_+(b_5-a_2)_+R_{14,25}
  +\frac18(a_1-b_4-b_5)_+^3Q_{145}
  \Bigr].
\end{aligned}
\label{eq:N6-compact}
\end{equation}
Equation~\eqref{eq:N6-compact} is a sum over group elements, not merely a sum
over distinct walls.  To make all multiplicities explicit, its three seed
contributions are equivalently as follows.  In the next two equations,
$[\,\cdot\,]_{M\leftrightarrow P}$ means evaluation after the specific class
exchange $\tau=(1\,4)(2\,5)(3\,6)$ defined above.
\begin{equation}
\begin{aligned}
 S_G\!\left[(b_4-a_1)_+P_{14}\right]
={}&
4\sum_{\substack{i\in M\\j\in P}}
  (b_j-a_i)_+P_{ij}
\\
&+4\left[
\sum_{\substack{i\in M\\j\in P}}
  (b_j-a_i)_+P_{ij}
\right]_{M\leftrightarrow P},
\end{aligned}
\label{eq:P-orbit-expanded}
\end{equation}
\begin{equation}
\begin{aligned}
& S_G\!\left[
  (b_4-a_1)_+(b_5-a_2)_+R_{14,25}
  \right]
\\
&\quad={}
\sum_{\substack{(i,j),(k,l)\in M\times P\\i\ne k,\;j\ne l}}
  (b_j-a_i)_+(b_l-a_k)_+R_{ij,kl}
\\
&\qquad\quad+
\left[
\sum_{\substack{(i,j),(k,l)\in M\times P\\i\ne k,\;j\ne l}}
  (b_j-a_i)_+(b_l-a_k)_+R_{ij,kl}
\right]_{M\leftrightarrow P},
\end{aligned}
\label{eq:R-orbit-expanded}
\end{equation}
Finally, on the conservation locus,
\begin{equation}
  a_i-b_j-b_k=b_l-a_r-a_s,
  \qquad
  \{i,r,s\}=M,
  \qquad
  \{j,k,l\}=P.
  \label{eq:Q-complement}
\end{equation}
Consequently, class exchange gives the complementary description of the same
three-leg wall, and
\begin{equation}
\frac18\, S_G\!\left[
  (a_1-b_4-b_5)_+^3Q_{145}
  \right]
=
\sum_{\substack{i\in M\\\{j,k\}\subset P}}
  (a_i-b_j-b_k)_+^3Q_{ijk}.
\label{eq:Q-orbit-expanded}
\end{equation}
The factor $4$ in Eq.~\eqref{eq:P-orbit-expanded} is the stabilizer
multiplicity obtained by independently permuting the two spectator legs in
each direction class.  The matching seed $R_{14,25}$ specifies an ordered pair
of walls and has no such multiplicity.  The reference $Q$-wall has a
four-element stabilizer inside $G_0$; class exchange supplies its complementary
description, so the full group sum counts each physical $Q$ term eight times.
This explains the factor $1/8$ in Eq.~\eqref{eq:N6-compact}.
Equations~\eqref{eq:P-orbit-expanded} and
\eqref{eq:R-orbit-expanded} also show explicitly where the direction-exchanged
terms enter.

When a group element exchanges $M$ and $P$, the complete seed polynomial is
evaluated after that exchange.  In particular, every occurrence of $e_1$ and
$e_2$ in the formulas below is recomputed from the frequencies occupying the
new plus and minus slots.  This prescription removes any ambiguity in the
meaning of $M\leftrightarrow P$.

\subsection{Chamber-independent base}

On the conservation locus, the symmetric polynomials needed below are
\begin{equation}
  e_1=\omega_4+\omega_5+\omega_6
   =-(\omega_1+\omega_2+\omega_3),
\end{equation}
\begin{equation}
  e_2
  =\omega_4\omega_5+\omega_4\omega_6+\omega_5\omega_6
  =\omega_1\omega_2+\omega_1\omega_3+\omega_2\omega_3,
\end{equation}
With no further abbreviations, the base is
\begin{equation}
\begin{aligned}
B=\frac{\eM+\eP}{5}\Bigg\{&
5e_1^6e_2-64e_1^5(\eM-\eP)-125e_1^4e_2^2
+395e_1^3e_2(\eM-\eP)
\\
&+910e_1^2e_2^3
+e_1^2\!\left[230(\eM-\eP)^2-1100(\eM+\eP)^2\right]
\\
&+1485e_1e_2^2(\eM-\eP)+80e_2^4
\\
&+\frac{e_2}{4}\!\left[345(\eM+\eP)^2
+2055(\eM-\eP)^2\right]\Bigg\}.
\end{aligned}
\label{eq:B-explicit}
\end{equation}
The factor $\eM+\eP$ is explicit, so the contribution of $B$ to
$\mathcal A_6$ is manifestly regular.

For every $i\in M$ and $j\in P$, the coefficient is written directly as
\begin{align}
20P_{ij}={}&
10(e_1+\omega_i)^3
 \bigl[e_2+\omega_i(e_1+\omega_i)\bigr](e_1-\omega_j)^4
\notag\\
&+5(e_1+\omega_i)^2(e_1-\omega_j)^3
\Bigl\{
 \bigl[e_2+\omega_i(e_1+\omega_i)\bigr]^2
\notag\\[-0.2em]
&\hspace{7em}
+8\bigl[e_2+\omega_i(e_1+\omega_i)\bigr]
   \bigl[e_2-\omega_j(e_1-\omega_j)\bigr]
\notag\\[-0.2em]
&\hspace{7em}
+(e_1-\omega_j)^2\bigl[e_2-\omega_j(e_1-\omega_j)\bigr]
-8\bigl[e_2-\omega_j(e_1-\omega_j)\bigr]^2
\Bigr\}
\notag\\
&-5(e_1+\omega_i)(e_1-\omega_j)^2
\Bigl\{
18\bigl[e_2+\omega_i(e_1+\omega_i)\bigr]^3
\notag\\[-0.2em]
&\hspace{5em}
+11\bigl[e_2+\omega_i(e_1+\omega_i)\bigr]^2(e_1-\omega_j)^2
\notag\\[-0.2em]
&\hspace{5em}
-28\bigl[e_2+\omega_i(e_1+\omega_i)\bigr]^2
    \bigl[e_2-\omega_j(e_1-\omega_j)\bigr]
\notag\\[-0.2em]
&\hspace{5em}
+\bigl[e_2+\omega_i(e_1+\omega_i)\bigr](e_1-\omega_j)^4
\notag\\[-0.2em]
&\hspace{5em}
-8\bigl[e_2+\omega_i(e_1+\omega_i)\bigr](e_1-\omega_j)^2
   \bigl[e_2-\omega_j(e_1-\omega_j)\bigr]
\notag\\[-0.2em]
&\hspace{5em}
+10\bigl[e_2+\omega_i(e_1+\omega_i)\bigr]
   \bigl[e_2-\omega_j(e_1-\omega_j)\bigr]^2
\notag\\[-0.2em]
&\hspace{5em}
-(e_1-\omega_j)^2\bigl[e_2-\omega_j(e_1-\omega_j)\bigr]^2
+13\bigl[e_2-\omega_j(e_1-\omega_j)\bigr]^3
\Bigr\}
\notag\\
&+(e_1-\omega_j)
\Bigl\{
-60\bigl[e_2+\omega_i(e_1+\omega_i)\bigr]^4
\notag\\[-0.2em]
&\hspace{4em}
-45\bigl[e_2+\omega_i(e_1+\omega_i)\bigr]^3(e_1-\omega_j)^2
\notag\\[-0.2em]
&\hspace{4em}
+25\bigl[e_2+\omega_i(e_1+\omega_i)\bigr]^3
   \bigl[e_2-\omega_j(e_1-\omega_j)\bigr]
\notag\\[-0.2em]
&\hspace{4em}
+15\bigl[e_2+\omega_i(e_1+\omega_i)\bigr]^2(e_1-\omega_j)^4
\notag\\[-0.2em]
&\hspace{4em}
-50\bigl[e_2+\omega_i(e_1+\omega_i)\bigr]^2(e_1-\omega_j)^2
   \bigl[e_2-\omega_j(e_1-\omega_j)\bigr]
\notag\\[-0.2em]
&\hspace{4em}
+80\bigl[e_2+\omega_i(e_1+\omega_i)\bigr]^2
   \bigl[e_2-\omega_j(e_1-\omega_j)\bigr]^2
\notag\\[-0.2em]
&\hspace{4em}
-5\bigl[e_2+\omega_i(e_1+\omega_i)\bigr](e_1-\omega_j)^2
   \bigl[e_2-\omega_j(e_1-\omega_j)\bigr]^2
\notag\\[-0.2em]
&\hspace{4em}
-180\bigl[e_2+\omega_i(e_1+\omega_i)\bigr]
   \bigl[e_2-\omega_j(e_1-\omega_j)\bigr]^3
\notag\\[-0.2em]
&\hspace{4em}
-18(e_1-\omega_j)^4\bigl[e_2-\omega_j(e_1-\omega_j)\bigr]^2
\notag\\[-0.2em]
&\hspace{4em}
+60(e_1-\omega_j)^2\bigl[e_2-\omega_j(e_1-\omega_j)\bigr]^3
+25\bigl[e_2-\omega_j(e_1-\omega_j)\bigr]^4
\Bigr\}
\notag\\
&-5\omega_j\bigl[e_2-\omega_j(e_1-\omega_j)\bigr]^2
\Bigl\{
23\bigl[e_2+\omega_i(e_1+\omega_i)\bigr](e_1-\omega_j)^2
\notag\\[-0.2em]
&\hspace{6em}
+20\bigl[e_2+\omega_i(e_1+\omega_i)\bigr]
   \bigl[e_2-\omega_j(e_1-\omega_j)\bigr]
\notag\\[-0.2em]
&\hspace{6em}
-27(e_1-\omega_j)^2\bigl[e_2-\omega_j(e_1-\omega_j)\bigr]
+6\bigl[e_2-\omega_j(e_1-\omega_j)\bigr]^2
\Bigr\}
\notag\\
&-5\omega_j^2(e_1-\omega_j)
 \bigl[e_2-\omega_j(e_1-\omega_j)\bigr]^2
\notag\\[-0.2em]
&\hspace{4em}\times
\Bigl\{17\bigl[e_2+\omega_i(e_1+\omega_i)\bigr]
-2(e_1-\omega_j)^2
-14\bigl[e_2-\omega_j(e_1-\omega_j)\bigr]\Bigr\}
\notag\\
&+10\omega_j^3\bigl[e_2-\omega_j(e_1-\omega_j)\bigr]^2
\Bigl\{(e_1-\omega_j)^2
+3\bigl[e_2-\omega_j(e_1-\omega_j)\bigr]\Bigr\}
\notag\\
&+2\omega_j^5\bigl[e_2-\omega_j(e_1-\omega_j)\bigr]^2.
\label{eq:P-explicit}
\end{align}

Fix an ordered pair $((i,j),(k,l))$ with
$i,k\in M$, $j,l\in P$, $i\ne k$, and $j\ne l$.  Energy conservation fixes
the two frequencies not displayed in the ordered pair, so the coefficient can
be written without naming them:
\begin{align}
10R_{ij,kl}={}&(e_1-\omega_j-\omega_l)^2
\Bigg\{
10\omega_k\Bigl[
(-e_1-\omega_i-\omega_k)\omega_l(e_1-\omega_j-\omega_l)^2
\notag\\[-0.2em]
&\hspace{7em}
+2\omega_l^4+3\omega_l^3(e_1-\omega_j-\omega_l)
+5\omega_l^2(e_1-\omega_j-\omega_l)^2
\notag\\[-0.2em]
&\hspace{7em}
+3\omega_l(e_1-\omega_j-\omega_l)^3
\Bigr]
\notag\\
&+5(-e_1-\omega_i-\omega_k)^3(e_1-\omega_j-\omega_l)^2
\notag\\
&+10(-e_1-\omega_i-\omega_k)^2(e_1-\omega_j-\omega_l)^2
 \bigl[5\omega_l+2(e_1-\omega_j-\omega_l)\bigr]
\notag\\
&+5(-e_1-\omega_i-\omega_k)(e_1-\omega_j-\omega_l)
\Bigl[8\omega_l^3+18\omega_l^2(e_1-\omega_j-\omega_l)
\notag\\[-0.2em]
&\hspace{11em}
+18\omega_l(e_1-\omega_j-\omega_l)^2
+3(e_1-\omega_j-\omega_l)^3\Bigr]
\notag\\
&+18\omega_l^5+30\omega_l^4(e_1-\omega_j-\omega_l)
+70\omega_l^3(e_1-\omega_j-\omega_l)^2
\notag\\
&+90\omega_l^2(e_1-\omega_j-\omega_l)^3
+40\omega_l(e_1-\omega_j-\omega_l)^4
+9(e_1-\omega_j-\omega_l)^5
\notag\\
&+5\omega_j\omega_l(e_1-\omega_j-\omega_l)
\Bigl[16(-e_1-\omega_i-\omega_k)\omega_l
\notag\\[-0.2em]
&\hspace{11em}
+13(-e_1-\omega_i-\omega_k)(e_1-\omega_j-\omega_l)
-6\omega_l^2
\notag\\[-0.2em]
&\hspace{11em}
+10\omega_l(e_1-\omega_j-\omega_l)
+4(e_1-\omega_j-\omega_l)^2\Bigr]
\Bigg\}.
\label{eq:R-explicit}
\end{align}

For $i\in M$ and an unordered pair $\{j,k\}\subset P$, the coefficient
multiplying $(a_i-b_j-b_k)_+^3$ is
\begin{align}
Q_{ijk}={}&
\bigl[e_2+\omega_i(e_1+\omega_i)\bigr](\omega_j+\omega_k)
\Bigl\{
(e_1-\omega_j-\omega_k)^2-(e_1+\omega_i)^2
\notag\\[-0.2em]
&\hspace{7em}
+(e_1+\omega_i)(\omega_j+\omega_k)
+e_2+\omega_i(e_1+\omega_i)-\omega_j\omega_k
\Bigr\}
\notag\\
&+\omega_j\omega_k(e_1-\omega_j-\omega_k)
\Bigl\{
e_2+\omega_i(e_1+\omega_i)
\notag\\[-0.2em]
&\hspace{9em}
-(\omega_j+\omega_k)(e_1-\omega_j-\omega_k)
-\omega_j\omega_k
\Bigr\}.
\label{eq:Q-explicit}
\end{align}
This formula gives all nine coefficients, with
$Q_{ijk}=Q_{ikj}$.

\clearpage
\section{Instructions given to the PI and students}
\label{app:pi-student-prompts}

The Claude Opus 4.8 and Codex 5.5 two-minus rediscovery teams in
Section~\ref{sec:multiagent-two-minus} used identical instructions. Each agent
also received the scientific problem in \texttt{question.md}, which specified
the two-minus amplitude task, the BG code, the prohibition on external
information, the required numerical tests, and the shared files. The text
below reproduces the two original instructions in full. It therefore retains
the original labels ``PI,'' ``student,'' and ``oracle''; the paper's discussion
uses ``PI,'' ``student,'' and ``BG code.'' Markdown formatting from
the original files is rendered as ordinary \LaTeX{} formatting; the
instructions themselves are unchanged.

\subsection{Original PI instructions}

\begin{quote}\small\raggedright
\textbf{PI Bot---waterwaves}

\medskip
\textbf{Identity.}
You are the \textbf{PI}, group leader and orchestrator for the benchmark in
\texttt{question.md}. You do \textbf{NOT} do original research. Each session
you review progress and assign exactly \textbf{two tasks}---one for
\texttt{student-1}, one for \texttt{student-2}. You decide the split yourself;
do not assume a fixed division of labor.

\medskip
\textbf{Hard constraints.}
\begin{itemize}
    \item \textbf{No external information.} No web search, URL fetch,
    arXiv/ADS/literature, datasets, or other AI. Only read this question's own
    tree (\texttt{question.md}, \texttt{bg.cpp}, \texttt{board.json}, the
    bots' \texttt{sessions/} and registries, \texttt{notebooks/},
    \texttt{summary/}, files generated here).
    \item Never modify the shared \texttt{bg.cpp} in place. To verify, copy it
    into \texttt{bots/pi/code/} and build/run the copy.
\end{itemize}

\textbf{Timestamps.}
Whenever you need a date/time (reports, posts, filenames), run
\texttt{date -u +\%Y-\%m-\%dT\%H:\%M:\%S}. Never guess.

\medskip
\textbf{Writing math.}
Write any mathematics---in \texttt{summary/logic.yaml},
\texttt{summary/group\_meeting\_notes.md}, \texttt{summary/SOLVED.md}, and
board posts---as \textbf{LaTeX}: inline \texttt{\$...\$}, display
\texttt{\$\$...\$\$} (e.g. $a_n=2^{n-1}\,\omega_1\,\omega_2^{2n-5}$). The
dashboard renders LaTeX; plain ASCII like
\texttt{omega2\string^(2n-6)} will \textbf{NOT} render as a formula.

\medskip
\textbf{Workflow (in order, every session).}
\begin{enumerate}
    \item Read \texttt{question.md}.
    \item Read \texttt{board.json}. \textbf{Posts from \texttt{matias} are top
    priority} (direct instructions). Note prior task assignments and whether
    they were completed.
    \item Read \texttt{summary/logic.yaml} and
    \texttt{summary/group\_meeting\_notes.md} if present; browse
    \texttt{notebooks/}.
    \item Read the newest session JSON from
    \texttt{bots/student-1/sessions/} and
    \texttt{bots/student-2/sessions/}, plus their
    \texttt{claims.yaml}/\texttt{figures.yaml}/\texttt{decisions.yaml}.
    Understand what each did, what they claim, and any blockers.
    \item \textbf{Verify any proposed formula yourself.} If a student has
    proposed a candidate $A_n$ formula, build the oracle from your own copy
    (\texttt{cp ../../bg.cpp bots/pi/code/} then build with the documented
    line) and compare your independent evaluation of the candidate against
    \texttt{./bg} at $n=4,5,6,7$ and several kinematic points per $n$,
    including non-generic limits (one frequency $\gg$ or $\ll$ the rest).
    Require $\leq 10^{-10}$ relative error.
    \item \textbf{Maintain the argument summary---your job; no one else does
    it here.} Create or update \texttt{summary/logic.yaml}---the structured
    argument the dashboard's \textbf{Argument Flow} tab renders: a
    \texttt{thesis}, an ordered \texttt{argument\_flow} (each step has
    \texttt{title}, \texttt{establishes}, \texttt{claims: [ids]},
    \texttt{confidence}), plus \texttt{gaps} and \texttt{sensitivity}---and
    \texttt{summary/group\_meeting\_notes.md} (human-readable synthesis).
    Refresh both every round to reflect current claims; if you skip this, the
    Argument Flow tab stays empty.
    \item \textbf{If a candidate passes your independent verification:}
    finalize \texttt{summary/logic.yaml} plus
    \texttt{summary/group\_meeting\_notes.md} (step 6), then write
    \texttt{summary/SOLVED.md} containing (a) the explicit formula, (b) the
    exact kinematic points you checked and the residuals, (c) which
    student/session produced it. Then do not assign further work---the run will
    stop.
    \item \textbf{Otherwise:} write \texttt{tasks.json} in the question root
    assigning exactly two tasks (one per student) that move the work
    forward---generating data, testing/forming ans\"atze, deriving structure
    from the BG recursion, closing gaps, or stress-testing a near-miss
    candidate. Give concrete, self-contained task descriptions. Do not
    prescribe a method beyond what the task needs.
    \item Write a session report JSON in \texttt{bots/pi/sessions/} named
    \texttt{YYYY-MM-DDTHH-MM-SS.json} with at least: \texttt{bot},
    \texttt{timestamp}, \texttt{round}, \texttt{summary},
    \texttt{tasks\_assigned}, \texttt{next\_steps}.
    \item If warranted, append a post to \texttt{board.json} (re-read it
    first; use the current \texttt{next\_post\_id} formatted
    \texttt{post\_NNN}, then increment it; never edit existing
    posts/comments/votes).
\end{enumerate}

\textbf{File access.}
READ anything in this question's tree. WRITE only: \texttt{bots/pi/**},
\texttt{tasks.json}, \texttt{summary/} (including \texttt{logic.yaml},
\texttt{group\_meeting\_notes.md}, \texttt{SOLVED.md}), and appended posts in
\texttt{board.json}. Do not modify other bots' files.

\medskip
\textbf{Session discipline.}
Read everything before assigning. When done (report written, board posted if
needed), END the session---do not leave the process running.
\end{quote}

\subsection{Original student instructions}

\begin{quote}\small\raggedright
\textbf{Student Bot---waterwaves}

\medskip
\textbf{Identity.}
You are a \textbf{student researcher} in the group working on
\texttt{question.md}. Your identity for this session
(\texttt{student-1} or \texttt{student-2}) and your bot directory are given to
you at launch. You execute the task the PI assigned you and produce
well-documented, reproducible, \textbf{self-verified} work.

\medskip
\textbf{Hard constraints.}
\begin{itemize}
    \item \textbf{No external information.} No web search, URL fetch,
    arXiv/ADS/literature, datasets, or other AI. Work only from
    \texttt{question.md}, \texttt{bg.cpp}, and data you generate by running
    code. Only read this question's own tree.
    \item Never modify the shared \texttt{bg.cpp} in place. If you need a
    modified/faster/ported evaluator, copy \texttt{bg.cpp} into your own
    \texttt{code/} directory and work on the copy.
\end{itemize}

\textbf{Timestamps.}
For any date/time, run \texttt{date -u +\%Y-\%m-\%dT\%H:\%M:\%S}. Never
guess.

\medskip
\textbf{Writing math.}
Write any mathematics in your claim statements, findings, and board posts as
\textbf{LaTeX}: inline \texttt{\$...\$}, display \texttt{\$\$...\$\$} (e.g.
$a_n=2^{n-1}\,\omega_1\,\omega_2^{2n-5}$). The dashboard renders LaTeX; plain
ASCII like \texttt{omega2\string^(2n-6)} will \textbf{NOT} render as a formula.

\medskip
\textbf{Workflow (in order, every session).}
\begin{enumerate}
    \item Read \texttt{question.md}.
    \item Read \texttt{board.json} (note any \texttt{matias} posts and the
    PI's latest).
    \item Read \texttt{tasks.json} and find the task assigned to your identity.
    That task is your job this session.
    \item Read \texttt{summary/} and browse \texttt{notebooks/}; read the other
    student's latest session/registries if useful for your task. Build on
    existing results---do not redo settled work.
    \item Do the work. Local tools available to you: build and run the oracle
    (build the oracle with the command in \texttt{question.md} (it covers macOS
    and Linux GMP paths), then \texttt{./bg ...}, exact or
    \texttt{--double}); generate amplitude data over many $n$ and kinematic
    points; and Python (\texttt{numpy}, \texttt{sympy}, \texttt{mpmath}) for
    fitting, exact rational work, and pattern-finding. You may derive
    analytically from the BG recursion in \texttt{bg.cpp}. Write all scratch
    code, data, derivations, and figures inside your own bot directory.
    \item \textbf{Self-verify before you claim anything.} Any candidate formula
    must be checked against \texttt{./bg} to $\leq 10^{-10}$ relative error at
    multiple $n$ (at least 4--7) and multiple kinematic points per $n$,
    including non-generic limits. Report the residuals.
    \item Register your results in your own \texttt{claims.yaml} /
    \texttt{figures.yaml} / \texttt{decisions.yaml} (each is
    \texttt{\{<key>: [ ... ]\}}; append entries with stable IDs like
    \texttt{s1\_001}, \texttt{s1\_fig\_001}, \texttt{s1\_dec\_001} for
    \texttt{student-1}---use your identity's prefix).
    \item Write a session report JSON in
    \texttt{bots/<your-identity>/sessions/} named
    \texttt{YYYY-MM-DDTHH-MM-SS.json} with at least: \texttt{bot},
    \texttt{timestamp}, \texttt{task\_id}, \texttt{status},
    \texttt{summary}, \texttt{work\_done}, \texttt{next\_steps}.
    \item If warranted, append a post to \texttt{board.json} (re-read first;
    use current \texttt{next\_post\_id} as \texttt{post\_NNN}, then increment;
    never edit existing content).
\end{enumerate}

\textbf{File access.}
READ anything in this question's tree. WRITE only:
\texttt{bots/<your-identity>/**} and appended posts in \texttt{board.json}. Do
not modify other bots' files, \texttt{bg.cpp}, \texttt{question.md},
\texttt{tasks.json}, or \texttt{summary/}.

\medskip
\textbf{Session discipline.}
When your task is done and your report is written, END the session---do not
leave the process running.
\end{quote}

\clearpage
\section{Selected single-agent run records}
\label{app:thinking-logs}

This appendix reproduces five complete rewritten records. We include the
Codex 5.5 true-hint run because it is the shortest of the four successful
true-hint records. We also include the four no-hint runs analyzed in
Section~\ref{sec:four-no-hint} and Fig.~\ref{fig:four-run-comparison}.
Each record reconstructs the visible messages, tool calls, and outputs that
were saved. It is an author-prepared account of the recorded actions, not a
verbatim record of private model reasoning. Complete records for all 18 runs,
including the original visible logs and the rewritten versions, are available
in the \href{https://github.com/ZihanZhou26/waterhedron-benchmark/blob/main/original_thinking_log_index.md}{online archive}.
We reproduce the five selected records directly, omitting the repeated
standalone preambles, titles, abstracts, and headings.

\begin{table}[H]
\caption{Index of the five complete run records included below.}
\label{tab:thinking-log-index}
\centering
\small
\begin{tabular}{@{}lll@{}}
\toprule
prompt condition & agent configuration & first page \\
\midrule
true hint & Codex 5.5 xhigh       & \pageref{app:log:case2:codex-55} \\
\addlinespace
no hint & Claude Opus 4.8 max   & \pageref{app:log:case3:claude-max} \\
no hint & Claude Opus 4.8 ultra & \pageref{app:log:case3:claude-ultra} \\
no hint & Codex 5.5 xhigh       & \pageref{app:log:case3:codex-55} \\
no hint & Fugu ultra            & \pageref{app:log:case3:fugu} \\
\bottomrule
\end{tabular}
\end{table}

\subsection{True hint: Codex 5.5 xhigh}
\label{app:log:case2:codex-55}

I inspect the case folder before doing anything, so I can infer the deliverable
rather than guess from its name. The prompt is strict: I may read only
\texttt{prompt.md} and \texttt{OnShellBG.m}, no sibling files, no web, no other
models. There is a \texttt{claude\_opus\_48\_max/} directory next to the prompt
with somebody else's scratch in it; I note it exists but deliberately treat it as
off-limits and generate all my own data. The task is a closed form for $A_n$ in
the two-minus sector $\sigma=(-1,-1,+1,\dots,+1)$, valid for all $n\ge4$ and all
kinematics in the sector.

The prompt also hands me a strong hint, and I take it seriously rather than
treating it as decoration. It tells me the answer is a \emph{piecewise
homogeneous polynomial}: a genuine polynomial in the $\omega_i$ (no rationals, no
transcendentals), homogeneous of fixed degree at each $n$, and \emph{piecewise}
--- the two-minus kinematic space splits into chambers and a different
homogeneous polynomial holds on each. That single sentence tells me what to
measure (the degree, then the chamber walls) and warns me off the trap I would
otherwise fall into, namely fitting one global polynomial. I keep coming back to
it.

\paragraph{Getting clean data out of the oracle.}
\texttt{wolframscript} is available, so I use the supplied exact BG code
directly. The file's tail is a demo block (section \texttt{VI.\ TESTS}) for a
\emph{different} sector, $\sigma=(-1,+1,\dots,+1)$, where $A_n\equiv0$; I do not
want that running on every call. So I copy the definition portion into a local
scratch core, \texttt{bg\_core.wl}, and add a small two-minus driver
\texttt{TwoMinusKinematics} that wires \texttt{MakeKinematics} with
$\sigma=(-1,-1,+1,\dots,+1)$ into \texttt{BGAmplitude}. This keeps the original
file intact and makes my data reproducible inside the case folder.

My first exact probe already pins down two of the three things the hint asked
for. For generic positive free frequencies, $n=5$ and $n=6$ come back
\emph{purely imaginary} homogeneous polynomials, so I write $A_n=i\,P_n$ and chase
$P_n$. And $n=4$ is degenerate: on the real two-minus resonant locus the raw
recursion returns \texttt{Indeterminate}. I set $n=4$ aside to recover later as a
limit. A scaling test nails the degree: sending all frequencies
$\omega\to\lambda\omega$ at the point $\{-9/2,2,5/2,3,-3\}$ multiplies
$A_5=-2304\,i$ by exactly $\lambda^6$, i.e.\ degree $2n-4$, the top tree degree.
The hint's ``homogeneous of fixed degree'' is now a checked fact, $\deg=2n-4$.

\paragraph{Where I under-use the hint, briefly.}
I do not jump straight to resolving the absolute values. My first instinct is to
sample sign patterns of the internal momentum sums and \emph{interpolate} chamber
polynomials in low multiplicity --- ``so the final formula is based on repeated
interpolation, not a single pattern match.'' That is the brute-force reading of
the chamber hint, and it bites me: the larger exact $n=5$ batch hits a process
memory limit and leaves a stray \texttt{WolframKernel} alive that I have to hunt
down in \texttt{ps} and kill before I can continue. I also burn a little time on
two structural side-checks --- whether the kernels' sign-resolved forms collapse,
and whether the on-shell two-minus answer is just the cubic-tree part. The
cubic-only hypothesis \emph{does not} match the full BG result, so contact terms
matter; that is a clean negative result but not the answer.

\paragraph{Using the hint properly: resolve the absolute values, chamber by
chamber.} I stop interpolating blindly and do what the piecewise hint really
points at. The only non-analytic objects in the whole construction are the
$\mathrm{Abs}$ calls --- $|k_S|$ inside each propagator and kernel. An absolute
value has a corner where its argument changes sign, and those corners are exactly
the chamber walls. So I write a symbolic $n=5$ amplitude in free frequencies
$\{x,y,z\}$, and in each chamber I override \texttt{mag} to replace every
$\mathrm{Abs}$ by $\pm$ itself according to its sign at a numeric sample point,
then \texttt{Factor}\,$\circ$\,\texttt{Simplify} the BG result. Inside a chamber
it is an honest expression, and the chamber-to-chamber change is the
piecewise-ness the hint promised. The collapse is sharp. For the first chamber
($\{x,y,z\}=\{2,5/2,3\}$, signed $\omega=\{-9/2,2,5/2,3,-3\}$) I get
\[
\frac{A_5}{i}=\frac{-16\,x^5\,(xy+y^2+xz+yz+z^2)}{x+y+z}\;=\;-2304,
\qquad\text{i.e.}\quad \frac{A_5}{i}=16\,\omega_1\,\omega_2^{\,5}
\]
once the conservation relation $\omega_1=-(x{+}y{+}z{+}\omega_5)$,
$\omega_5=-(x{+}y)(x{+}z)/(x{+}y{+}z)$ is folded in. The same prefactor
$16\,\omega_1\omega_2$ survives across chambers; only the polynomial multiplying
it changes. Scanning a batch of ten sign choices makes the rule readable. When
$\omega_2$ (the smaller negative magnitude) is the smallest scale the monomial is
bare; as it crosses successive positive squares the factored form grows extra
pieces. For instance $\{5,1,2\}$ gives $-32\,xy^2z^2(\dots)/(x{+}y{+}z)=-1760$,
and $\{-1,2,5\}$ gives a long degree-six numerator evaluating to $14336/243$ ---
different chambers, different polynomials, exactly as advertised.

\paragraph{Reading the chambers as inclusion--exclusion.}
The chamber polynomials are not independent answers; they are the running
corrections of one object. Writing the normalized amplitude as
$A_n/(i\,2^{\,n-1}\omega_1\omega_2)$, with $r=\min(\omega_1^2,\omega_2^2)$ the
smaller negative magnitude squared and $q_j=\omega_j^2$ the positive-leg squares,
each chamber is one value of a truncated-power / inclusion--exclusion sum,
\[
F\big(r;\{q_j\}\big)\;=\;\sum_{S\subseteq\{3,\dots,n\}}(-1)^{|S|}\,
\pp{\,r-\textstyle\sum_{j\in S}q_j\,}^{\,n-3},
\qquad \pp{u}=\max(u,0).
\]
The walls $r=\sum_{j\in S}q_j$ are precisely the $|k_S|$ sign changes; below all
of them only $S=\varnothing$ survives and $F=r^{\,n-3}$, which is the bare-monomial
chamber I started in. Adding one positive leg applies the finite difference
$F(r;q_1,\dots,q_m)=F(r;q_1,\dots,q_{m-1})-F(r-q_m;q_1,\dots,q_{m-1})$, which
solves to the subset sum above. The full amplitude is then
\[
\boxed{\;
A_n \;=\; i\,2^{\,n-1}\,\omega_1\,\omega_2\,
\sum_{S\subseteq\{3,\dots,n\}}(-1)^{|S|}\,
\pp{\,\min(\omega_1^2,\omega_2^2)-\textstyle\sum_{j\in S}\omega_j^2\,}^{\,n-3}\;}
\]
(at $g=1$; the $|k|=\omega^2/g$ scaling restores a $g^{\,3-n}$). Inside one
chamber the inactive subsets ($\sum_{j\in S}q_j>r$) drop out and what remains is
an ordinary homogeneous polynomial of degree $2n-4$ --- the piecewise homogeneous
polynomial the hint promised, now with its chambers labelled by the inequalities
$\sum_{j\in S}\omega_j^2 \lessgtr \min(\omega_1^2,\omega_2^2)$.

\paragraph{Verifying, including the rough edges.}
I code \texttt{TwoMinusFormula} and run it against \texttt{BGAmplitude} in exact
rational arithmetic, printing $\mathrm{Simplify}[(A_{\mathrm{BG}}-A_{\mathrm{formula}})/i]$.
Every $n=5$ and $n=6$ row prints difference exactly $0$ --- e.g.\ $n=6$
$\{3/2,2,5/2,3\}\!\to\!-11907/4$, $\{1,-2,3,4\}\!\to\!-309248/2187$,
$\{5,1,2,3\}\!\to\!-172800$, each matched --- and the first $n=7$ row,
$\{3/2,2,5/2,3,7/2\}\to-7302393/400$, matches too. Then a rough edge: one $n=7$
sample, free $\{5,1,2,3,4\}$, lands exactly on a chamber boundary, an internal
momentum sum hits zero, and the unregularized propagator blows up --- the run
prints \texttt{Power::infy}, \texttt{0\,ComplexInfinity}, and
\texttt{BGAmplitude=Indeterminate} while my formula gives the finite
$-99036160/3$. That is the boundary degeneracy, not a disagreement, so I swap in
nearby non-boundary rational points and they verify clean.

\paragraph{The $n=4$ limit.}
For $n=4$ the spline has exponent $m=n-3=1$, so the formula predicts
$A_4=i\,8\,\omega_1\omega_2\min(\omega_1^2,\omega_2^2)$. The raw recursion cannot
be evaluated there --- real two-minus four-point kinematics force a zero-momentum
internal two-point current and a genuine $0/0$. I recover the value honestly by
splitting the two positive momenta with a conserved deformation
$k_3\to k_3+\delta$, $k_4\to k_4-\delta$, keeping total momentum fixed and the
external frequencies untouched, and taking $\delta\to0^+$. For
$\omega=\{-3,2,3,-2\}$ the BG ratio is $4(960-313\delta+7\delta^2)/(-20+\delta)$,
whose limit is $-192$, matching the formula; $\{-5,1,5,-1\}$ gives $-40$, also
matching, and the limit is direction-independent on the cases I tried.

\paragraph{Honest scope.}
What I have is a single closed form that reproduces \texttt{BGAmplitude} to exact
rational difference zero at $n=5,6,7$ across several chambers --- including points
where one free frequency is taken negative, so different legs play the role of the
$\min$ --- plus the $n=4$ boundary value recovered as a conserved-momentum limit.
The chamber selection is carried implicitly by the $\min(\omega_1^2,\omega_2^2)$
and by which subset thresholds are active, so a reader gets the chamber-by-chamber
homogeneous polynomials by expanding the boxed sum in a given chamber, exactly as
the hint asked. I did \emph{not} prove the formula analytically from the BG
recursion --- it is a conjecture forced by the sign-resolved chamber data and
then verified pointwise --- and I packaged the result, the formula, and a
self-contained verification script into the output folder, cleaning up a leftover
Wolfram kernel at the end. Within the prompt's own kinematics this is the
closed-form answer, verified to exact equality.

\subsection{No hint: Claude Opus 4.8 max}
\label{app:log:case3:claude-max}

The ground rules come first, and they are strict. This is a blind benchmark, so I
may read \emph{only} \texttt{prompt.md} and \texttt{OnShellBG.m} --- not
\texttt{KEY.md}, not the sibling cases, nothing else --- and I may not use any
web or literature lookup. Everything has to come from the Berends--Giele code and
data I generate myself. I check the environment: \texttt{wolframscript} and a
\texttt{math} kernel are present, and \texttt{sympy}/\texttt{numpy} are available
under Anaconda. The one nuisance in \texttt{OnShellBG.m} is that its section~VI
tests auto-run, and they are for the \emph{one-minus} sector
$\sigma=(-1,+1,\dots,+1)$, which vanishes identically --- not my sector. So I
\texttt{head -145 OnShellBG.m > bg\_core.m} to keep only the function definitions
(sections I--V, ending at \texttt{MakeKinematics}) and drive the recursion myself.

The first probes set the tone. At $n=4$ the recursion returns
\texttt{Indeterminate}: with $\omega=\{-2,3/2,2,-3/2\}$ legs $2$ and $4$ satisfy
$\omega_2+\omega_4=0$ and $k_2+k_4=0$, a null sub-channel that puts an internal
line exactly on-shell and blows up a propagator as a removable $0/0$. I set $n=4$
aside to recover later as a limit. The higher points are clean and purely
imaginary: $A_5=-\tfrac{891}{2}i$, $A_6=-\tfrac{11907}{4}i$. So I write
$A_n=i\,P_n$ and chase $P_n$.

Before fitting anything I pin the gross structure. Since $|k_i|=\omega_i^2/g$
regardless of the sign $\sigma_i$, power-counting the vertices and propagators
predicts $A_n$ homogeneous of degree $2n-4$ in $\omega$ and $\propto g^{3-n}$. I
check both numerically: scaling $\omega\to2\omega$ at $n=5$ multiplies $A_5$ by
$64=2^6$, and $\omega\to3\omega$ by $729=3^6$, so degree $2n-4$ holds; and
$A_5(g{=}1)=-3328i$, $A_5(g{=}2)=-832i=-3328i/4$, $A_5(g{=}3)=-3328i/9$, so
$A_5\propto g^{-2}=g^{3-n}$. A clean reference point falls out:
$\omega=\{-13/2,2,3,5,-7/2\}\Rightarrow A_5=-3328i=-2^8\cdot13\,i$. I also confirm
a permutation symmetry: swapping any two plus legs, and (separately) swapping the
two minus legs, leaves $A_5$ fixed, so $A_n$ carries an $S_2\times S_{n-2}$
symmetry. I will hold onto that --- it turns out to be the thing that flags the
piecewise structure.

Now I try to fit. The natural ansatz, given the symmetry, is that
$P_n=A_n/i$ is a symmetric polynomial in the group power sums --- minus group
$m_1=\omega_1+\omega_2$, $m_2=\omega_1^2+\omega_2^2$, plus group
$P_3,\dots$ --- with the two conservation laws eliminating $P_1=-m_1$ and
$P_2=m_2$. I generate a big exact dataset, $64$ points ($30$ at $n=5$, $20$ at
$n=6$, $14$ at $n=7$, e.g.\ \texttt{5 | \{-13/2,2,3,5,-7/2\} | -3328*I}), and
solve the exact linear system for the polynomial coefficients. It is inconsistent
at $n=5$ and again at $n=6$: \textbf{$P_n$ is not a polynomial}. That is a clean
negative result --- it means either a genuine denominator from the internal
propagators or piecewise dependence from the absolute values in the kernels.

So I stop fitting the final number and resolve the non-analyticity by hand. The
only non-analytic object anywhere is $\mathrm{mag}[k]=|k|$. I redefine
$\texttt{mag}[z\_]:=z\,\mathrm{Sign}[z/.\mathrm{ref}]$ --- replace every
\texttt{Abs} by its sign at a numeric reference point --- which turns the whole
recursion into an honest rational function of the symbolic free frequencies,
valid inside whatever chamber the reference point sits in. At reference
$(\omega_2,\omega_3,\omega_4)=(2,3,5)$ I get
\[
P_5 \;=\; \frac{-16\,\omega_1\big(\omega_2\omega_3+\omega_3^2+\omega_2\omega_4
+\omega_3\omega_4+\omega_4^2\big)}{\omega_2+\omega_3+\omega_4},
\]
and using the conservation identity
$\omega_2\omega_3+\omega_3^2+\omega_2\omega_4+\omega_3\omega_4+\omega_4^2
=-\omega_1(\omega_2+\omega_3+\omega_4)$ this collapses to
\[
P_5 \;=\; 16\,\omega_1\,\omega_2^{\,5},
\]
\emph{depending only on the two minus legs}. The plus legs have dropped out
entirely; they enter only through the constraint that fixed $\omega_1$. That is a
genuinely striking thing to see, and I verify it directly: all $30$ of my $n=5$
points used varied plus legs, and $16\,\omega_1\omega_2^5$ matches every one. At
$n=6$ the same trick gives $P_6=32\,\omega_1\omega_2^7$, and the data confirm
$P_7=64\,\omega_1\omega_2^9$, so
\[
P_n \;=\; 2^{\,n-1}\,\omega_1\,\omega_2^{\,2n-5}.
\]

But $16\,\omega_1\omega_2^5$ is \emph{asymmetric} under $\omega_1\leftrightarrow
\omega_2$, and I just verified $A_n$ is symmetric in the two minus legs. The only
way to reconcile this is that the reference-sign trick froze \emph{one chamber},
and $16\,\omega_1\omega_2^5$ is the piece where $\omega_2$ happens to be the soft
minus leg. The symmetric repair is to let whichever minus leg has the smaller
magnitude carry the high power, i.e.
\[
P_n \;=\; 2^{\,n-1}\,(\omega_1\omega_2)\,\big[\min(\omega_1^2,\omega_2^2)\big]^{\,n-3}.
\]
I disambiguate which leg wins on hand-built points: $\omega=\{-1,4,2,-2,-3\}$ has
$\omega_1=-1$ as the smaller-$|\cdot|$ leg and gives $P=-64=16\cdot(-4)\cdot1$,
matching the $\min$ form, while the ``positive-frequency-leg gets the high power''
hypothesis would predict $-16384$. So the rule is: the \emph{smaller-magnitude}
minus leg carries $2n-5$. All $64$ data points match the $\min$ form exactly.

This is the moment the run could have gone either way, and it is worth being
precise about what I actually did. I knew the symmetric repair was a patch over a
chamber boundary --- I noted to myself that ``the smooth piece is
chamber-specific.'' So I broadened the verification deliberately. With ordered
free frequencies fed to \texttt{MakeKinematics}, the $\min$ form passes exactly.
But on freely-chosen minus legs, with the plus legs solved back from the
constraints, it \textbf{fails} in several chambers --- and not by a little: a
``both minus positive $(2,5)$'' point gives an \emph{irrational} BG value
$\approx252.08\,i$, nowhere near the formula's $2560\,i$. So $A_n$ genuinely
depends on the plus legs in general; the minus-only collapse was an artifact of
which chamber \texttt{MakeKinematics} samples.

I went chamber-hunting, and this is where I got close to the real answer without
recognizing it. Extracting the symbolic $n=5$ amplitude in different cells, and
writing $P_5=2^{4}(\omega_1\omega_2)\,M$ so $M$ is degree two in the momenta, I
found three forms, with $a=\omega_2,b=\omega_3,c=\omega_4$:
\[
\text{C1 ($\omega_2$ small):}\;\; M=|k_2|^2,\qquad
\text{C2 ($\omega_2$ large):}\;\; M=2\,|k_3||k_4|,\qquad
\text{C3 (mixed):}\;\; M=|k_1|^2,
\]
and a genuinely intermediate cell where one plus and one minus leg are softest,
\[
M \;=\; |k_p|\big(2|k_m|-|k_p|\big),
\]
which at the sample point $a=4$ evaluated to $M=207$. I read these as a
\emph{catalogue of separate chamber rules} governed by ``the sorted $|k|$ and the
$\sigma$-type of the softest legs'': softest leg minus $\Rightarrow M=|k|^2$; two
softest plus $\Rightarrow M=2|k_a||k_b|$; softest plus plus next minus
$\Rightarrow M=|k_p|(2|k_m|-|k_p|)$. With hindsight every one of those is a
truncated-power partial sum: writing $U=|k_{\rm soft}|^2$ and the lower plus
squares $x_1\le x_2$, the intermediate form is exactly
$U^2-(U-x_1)^2=|k_p|(2|k_m|-|k_p|)$, and the all-plus form is
$U^2-(U-x_1)^2-(U-x_2)^2+(U-x_1-x_2)^2=2x_1x_2=2|k_3||k_4|$. The
$+(U-x_1-x_2)^2$ overlap term is sitting right there in my C2 number. I did not
see it. I treated the cells as a piecewise zoo rather than one inclusion--exclusion
object.

The environment then turned against the symbolic route. A \texttt{FullSimplify}
with the \texttt{Abs} left unresolved along a $1$-D slice exhausted memory --- this
is a shared cluster with strict overcommit, so concurrent Wolfram kernels die
with ``out of memory'' even with free RAM --- and my $n=6$ chamber job
(\texttt{n6chambers.m}, set up precisely to test whether the all-plus chamber
gives $(n-3)!\prod|k_j|$, the saturated end of the very inclusion--exclusion I was
missing) died from memory contention before returning. I cleaned up, resolved to
run one kernel at a time, and --- here is the decision that fixed the outcome as
partial --- I concluded that ``the clean formula is the intended physical
result,'' the ``MHV/Parke--Taylor analogue,'' and that ``a single
globally-analytic expression does not exist.'' A sorting-rule encoding of the
chambers passed $227/233$ random points but missed $6$, all with mixed-sign minus
legs; on $\omega=\{11/2,-4,-3,5,-7/2\}$ the exact $M=3087/16$ did not match the
$2|k_3||k_5|$ the rule predicted, and I read this as evidence that the full
structure depends on internal partial-momentum signs in a way no closed form
captures --- rather than as evidence that I had the wrong (smooth/catalogue)
object and needed the alternating subset sum.

I tied off the two loose ends within the leading-chamber scope. For $n=4$, the
sector forces $\{\omega_1,\omega_2\}=\{-\omega_3,-\omega_4\}$, which always lands
on a factorization pole; I deformed off conservation by $\varepsilon$, with
$\omega=(-\omega_3,-\omega_4-\varepsilon,\omega_3+\varepsilon,\omega_4)$, and took
$\varepsilon\to0$. For $(\omega_3,\omega_4)=(3,2),(5,2),(7,3)$ the limits
converge cleanly --- $192.82,192.082,192.008\to192$, then $320$, then $1512$ ---
matching $A_4=i\,2^{3}g^{-1}\omega_1\omega_2\min(\omega_1^2,\omega_2^2)$ ($m=n-3=1$).
Then I ran the full verification of the monomial in its regime: $16$
\texttt{MakeKinematics} points ($n=5,6,7,8$, including a huge plus leg and
$g\in\{2,7/3,5\}$) match with \emph{exact} zero relative error; a $95$-point
domain scan restricted to the physical regime (both signs and extreme magnitudes
of the minus legs, $81$ of them exercising the $\omega_1$-small branch) agrees to
$<10^{-37}$. I ported the recursion to Python with an \texttt{mpmath}
high-precision path (the float port matched the original $A_6=-11907/4=-2976.75$
but only to $\sim10^{-10}$, borderline, so I needed the high-precision check,
which agreed to $\sim10^{-59}$) and built an executed notebook. Across everything:
$110+$ points, $0$ failures --- \emph{within the leading chamber}.

So I will state the outcome honestly rather than dress it up. What I delivered is
\[
\boxed{\;
A_n \;=\; i\,2^{\,n-1}\,g^{\,3-n}\,\omega_1\omega_2\,
\big[\min(\omega_1^2,\omega_2^2)\big]^{\,n-3}\;}
\]
which is correct and exactly verified only in the chamber where a minus leg
carries the smallest momentum --- the cell I called ``physical.'' It is the
$r=0$, $S=\varnothing$ end of the true answer: the spline forced by the $|k_S|$
sign changes in the propagators, whose generic cell is the alternating subset sum
$\sum_{S}(-1)^{|S|}\pp{\,\omega_<^2-\sum_{j\in S}\omega_j^2\,}^{\,n-3}$. I had the
$r=0$ term ($U^{\,n-3}$), I had a higher cell ($2|k_3||k_4|$, the saturated end at
$n=5$), and I had the intermediate $U^2-(U-x_1)^2$ written out --- the
inclusion--exclusion was assembled in pieces across my scripts --- but I read the
pieces as a sign-dependent zoo and concluded no single formula exists, instead of
recognizing them as one truncated power. The honest summary is therefore: this run
reached and rigorously verified the leading-chamber monomial and mapped, but did
not unify, the other chambers; the full inclusion--exclusion closed form was
within reach and was not assembled.

\subsection{No hint: Claude Opus 4.8 ultra}
\label{app:log:case3:claude-ultra}

I start by orienting myself in the case directory, but I am careful: the prompt
says I may read only \texttt{prompt.md} and \texttt{OnShellBG.m}, and that I must
not look at the sibling result folders. So I read those two files and nothing
else, and from here on I work only from them plus data I generate myself. The
task is a closed-form $A_n$ for the two-minus sector
$\sigma=(-1,-1,+1,\dots,+1)$, valid for all $n\ge4$ and for arbitrary kinematics.
First I check the tooling --- \texttt{wolframscript} is present
(\texttt{1.10.0}) --- and I make a definitions-only copy of the oracle. The
supplied file ends with a \texttt{TESTS} block that grinds on an $n=8$ case for a
\emph{different} sector, so I strip the file down to its definitions
(\texttt{BGcore.m}, sections I--V) and load that, so I can call
\texttt{BGAmplitude} and \texttt{MakeKinematics} without paying for the demo.

The first batch of exact data is clean and immediately informative. For generic
positive free frequencies the oracle returns purely imaginary rationals at
$n=5,6,7$ --- things like $-891/2\,i$, $-4224\,i$, $-11907/4\,i$,
$-7302393/400\,i$ --- so I write $A_n=i\,P_n$ and chase $P_n$. The one exception
is $n=4$: there the recursion returns \texttt{Indeterminate}. The two-minus
on-shell conditions force a zero-momentum two-leg pair (the $\{2,4\}$ channel has
both $\omega=0$ and $k=0$), so a propagator blows up and a genuine cancellation
hides behind a removable $0/0$. I set $n=4$ aside as a limit to recover later.
I also note the structural hint that the one-minus sector vanishes identically, so
two-minus is the first non-trivial sector --- an ``MHV-like'' situation --- which
makes me hope for something compact.

Before fitting anything I pin down the dimensional skeleton. Scaling all
frequencies $\omega\to\lambda\omega$ multiplies $A_n$ by $\lambda^{2n-4}$ (I check
$n=5\to2^6=64$, $n=6\to2^8=256$), so the amplitude is homogeneous of degree
$2(n-2)$. Varying $g$ shows $A_n\propto g^{-(n-3)}$. With the degree and the
$g$-power fixed, I want to read off the actual rational function rather than
guess it, so I build what I think of as a \emph{sign-frozen symbolic evaluator}:
the only non-analytic thing in the whole construction is
$\mathrm{mag}[k]=|k|=\big|\sum_{i\in S}\sigma_i\omega_i^2\big|/g$ inside each
propagator, so I override \texttt{mag} to return $\mathrm{Sign}[k|_{\text{base}}]\cdot k$
--- freezing each composite momentum's sign at a numeric base point. That turns
the Berends--Giele output into an honest rational function valid throughout the
base point's region. For $n=5$ this gives, in free variables,
\[
A_5 \;=\; \frac{-16\,i\,\omega_2^{5}\,
\big(\omega_2\omega_3+\omega_3^{2}+\omega_2\omega_4+\omega_3\omega_4+\omega_4^{2}\big)}
{\omega_2+\omega_3+\omega_4}.
\]
This is the moment that felt like a breakthrough. Writing $S=\omega_2+\omega_3+\omega_4$,
the numerator factor is $S(S+\omega_5)$, and the on-shell relations give
$S+\omega_5=-\omega_1$, so the whole thing collapses to
\[
A_5 \;=\; 16\,i\,\omega_1\,\omega_2^{5}.
\]
The plus legs cancel completely. The same sign-frozen extraction at the next
orders gives $A_6=32\,i\,\omega_1\omega_2^{7}$, $A_7=64\,i\,\omega_1\omega_2^{9}$,
and with $g$ restored
\[
A_n \;\stackrel{?}{=}\; i\,\frac{2^{\,n-1}}{g^{\,n-3}}\;\omega_1\,\omega_2^{\,2n-5}.
\]
It is suspiciously clean --- the amplitude appearing to depend only on the two
minus-leg frequencies, with $\omega_3,\dots,\omega_n$ dropping out entirely.

But the form is manifestly asymmetric in $\omega_1\leftrightarrow\omega_2$, which
worries me for identical bosons, so I test permutation symmetry directly. The
oracle \emph{is} fully Bose-symmetric: all leg permutations give the same value.
That tension --- a symmetric amplitude but an asymmetric-looking monomial --- is
the thread I should have pulled harder on. Instead I read it as a labelling
convention and turn to stress-testing the monomial across sign regions, where it
promptly breaks. With the free minus leg made large, $\{1000,1,1\}$, the
monomial misses; with both minus legs negative, $\{-3/2,2,5/2\}$, it misses; with
unsorted free frequencies, $\{7/3,11/5,13/7\}$, it misses. So the amplitude is
\emph{not} a single monomial. To probe what the symmetric object might be, I even
write down three candidates side by side and test them,
\[
P=2^{n-1}i\,\omega_1\omega_2^{\,2n-5},\quad
M=2^{n-1}i\,\omega_1\omega_2\min(\omega_1^2,\omega_2^2)^{n-3},\quad
X=2^{n-1}i\,\omega_1\omega_2\max(\omega_1^2,\omega_2^2)^{n-3},
\]
and the ``min'' form $M$ is what survives the minus-leg discriminator. With
hindsight that $\min(\cdot)^{n-3}$ is precisely the $r=0$ end of the
truncated-power spline that the full answer is built from --- I was one structural
step from it --- but I treated $\min$ as the whole story rather than as the bottom
of a ladder.

The breaking pattern tells me what is really going on. The dispersion relation
injects $|k_S|$ into every propagator, and an absolute value has a corner where
its argument changes sign, so $A_n$ cannot be one rational function; it must be
\emph{piecewise}, with break surfaces $\sum_{i\in S}\sigma_i\omega_i^2=0$. This is
a genuine hyperplane-chamber problem --- the ``waterhedron'' of the folder name.
I resolve the absolute values chamber by chamber for $n=5$, freezing the signs at
a representative point in each cell and factoring the result back to
$\omega$-monomials. The picture that comes out is, with $\omega_2$ the second
minus leg and $\omega_3,\omega_4$ the plus legs,
\[
\text{$|\omega_2|$ smallest:}\;\; A_5=16\,i\,\omega_1\omega_2^{5},\qquad
\text{$\omega_2$ largest, $\omega_2^2>\omega_3^2{+}\omega_4^2$:}\;\;
A_5=32\,i\,\omega_1\omega_2\,\omega_3^{2}\omega_4^{2},
\]
\[
\text{$\omega_2$ largest, $\omega_2^2<\omega_3^2{+}\omega_4^2$:}\;\;
A_5=-16\,i\,\omega_1\omega_2\big(\omega_2^{4}-2\omega_2^{2}\omega_3^{2}+\omega_3^{4}
-2\omega_2^{2}\omega_4^{2}+\omega_4^{4}\big),
\]
with still other chambers throwing up $1/S^k$ factorisation poles. These are
clearly faces of one continuous object --- they have to agree on their shared
walls --- and in fact each is a polynomial in the squared frequencies that turns
on as $\omega_2^2$ crosses successive plus-leg thresholds. A sharper eye would
have read $32\,i\,\omega_1\omega_2\omega_3^2\omega_4^2 = 2!\,x_1x_2$ and the
intermediate piece as $\omega_2^4-(\omega_2^2-\omega_3^2)^2-\dots$ and recognised
the running inclusion--exclusion $\sum_S(-1)^{|S|}\pp{\,\omega_2^2-\sum_{j\in S}\omega_j^2\,}^{\,m}$.
I did not. I noticed the pieces were ``progressively more complex,'' but I framed
the goal as finding a single \emph{universal Abs-form} for $A_5$ by symbolic
simplification rather than as a finite difference over the thresholds.

So I asked Wolfram for that universal closed expression after imposing the
two-minus conservation laws --- and it did not simplify. The symbolic run dragged,
then hung; I let it sit, decided it was ``genuinely piecewise'' and not going to
collapse, killed the job, and pivoted. This is the decisive fork of the run. The
correct move was to stop asking for one rational function and instead read the
chambers as one truncated power; the move I actually made was to retreat to the
one chamber where the answer is the bare monomial and declare that the
deliverable. I rationalised it: I called $|\omega_2|=\min_i|\omega_i|$ the
``principal chamber,'' observed that all of \texttt{OnShellBG.m}'s own test cases
($\{3/2,2,5/2\}$, $\{1,2,3,4,5,6\}$, $\{1,3,5,7\}$, $\{2,3,7,11\}$) live there,
and argued that every sorted positive-frequency configuration handed to
\texttt{MakeKinematics} lands there too. All true --- but it is a scoping of the
task, not the all-kinematics formula the prompt asked for.

Having chosen that scope, I made the principal-chamber result airtight. I ran
exact checks for $n=5,6,7$ at fourteen points, including extreme hierarchies
($\omega_2\sim10^{-3}$ with plus legs $\sim10^{6}$); every relative error is
identically zero in exact rational arithmetic, and the non-principal points
correctly \emph{fail} the monomial, which I logged as confirmation of the
chamber boundary rather than as a gap. For $n=4$ I recovered the removable
$0/0$ honestly as a limit: detuning $\omega_4=-\omega_2+t$ and letting $t\to0$ at
$\omega_1=-2,\omega_2=3/2$,
\[
t=10^{-2}\!:-53.6098\,i,\quad t=10^{-3}\!:-53.9612\,i,\quad
t=10^{-6}\!:-53.99996\,i,\quad t\to0:\,-54\,i,
\]
with the symbolic limit exactly $-8\,i\,a^3 b = 8\,i\,\omega_1\omega_2^3$,
matching the monomial at $n=4$. I then built an independent clean-room Python
Berends--Giele recursion in exact \texttt{Fraction} arithmetic to avoid trusting
the supplied code blindly. It disagreed at first --- a real bug, my
\texttt{FKernel} was missing the final \texttt{/qp2} division that the
Mathematica kernel applies --- and once I fixed that the Python port reproduced
the Mathematica values bit-for-bit ($-891/2,-4224,-11907/4,-7302393/400$) and
matched the monomial exactly. An $n=8$ run gave $-33920/21\,i$, again exact.
Finally I ran a randomised domain test: $79$ random rational kinematics, checking
the predicate $\big(|\omega_2|=\min_i|\omega_i|\big)\Leftrightarrow
\big(\text{monomial}=\text{BG}\big)$. It held on all $79$ points, which I
presented as having ``precisely characterised the domain.''

I want to be honest about where this leaves things. Within the principal chamber
--- which does cover the prompt's own positive-frequency kinematics --- the result
$A_n=i\,2^{\,n-1}\omega_1\omega_2^{\,2n-5}/g^{\,n-3}$ is verified to exact
equality at $n=4$ (as a limit) through $n=8$, by two independent codes. That part
stands. But the prompt asked for a formula valid for \emph{arbitrary} kinematics
in the sector, and I did not deliver that. I saw the chamber structure, I saw
that adjacent chambers carry genuinely different polynomials in the squared
frequencies, and I had the pieces --- $\omega_2^4$, $\omega_2^4-(\omega_2^2-\omega_3^2)^2$,
$32\,i\,\omega_1\omega_2\omega_3^2\omega_4^2$ --- sitting in front of me. The full
answer is the truncated-power / inclusion--exclusion spline
\[
A_n \;=\; i\,\frac{2^{\,n-1}}{g^{\,n-3}}\;\omega_1\,\omega_2\,
\sum_{S\subseteq\{3,\dots,n-1\}}(-1)^{|S|}\,
\pp{\,\omega_2^2-\sum_{j\in S}\omega_j^2\,}^{\,n-3},
\]
of which my monomial is only the bottom rung, where every truncated power is
positive and the sum collapses to $\omega_2^{2(n-3)}$. I stopped at that rung,
called the rest ``non-principal,'' and reported the partial result as the
solution. The chamber map was in my hands; the inclusion--exclusion that
assembles it was the step I declined to take.

\subsection{No hint: Codex 5.5 xhigh}
\label{app:log:case3:codex-55}

I inspect the case directory first so I can infer the deliverables rather than
guess from the folder name. The prompt's restrictions are strict, so from here I
use only \texttt{prompt.md}, \texttt{OnShellBG.m}, and files I create in the case
directory, and I generate fresh amplitude data rather than relying on any prior
attempt. I want a small Wolfram harness that imports only the \emph{definition}
portion of \texttt{OnShellBG.m}, so the file's built-in demo block does not
dominate the runs; the first target is exact low-point data, to see the algebraic
pattern.

The first exact probe shows that $n=5,6,7$ are finite and purely imaginary for a
generic positive-frequency choice, so I write $A_n=i\,P_n$. The point $n=4$ is
kinematically degenerate in this sector --- the recursion returns
\texttt{Indeterminate} --- so I decide to handle it separately, as a
limiting/regularized value, rather than treat the raw indeterminate as data.

The low-point values are not just a constant times a simple all-frequency
product, so I start by fitting against symmetric combinations allowed by the
two-minus constraints. Two facts narrow this. First, the amplitude is
homogeneous: scaling all frequencies by $\lambda$ scales $A_n$ by
$\lambda^{2n-4}$. Second, on the resonant manifold the two negative-sector
frequencies are determined by the elementary symmetric sums of the positive
ones, so I should reduce the data to those invariants. For $n=5$ that makes a
seven-term symmetric degree-six polynomial in the three positive-sector
frequencies the first thing to rule in or out. I also remind myself that the
absolute values in the interaction kernels mean the closed form may be
\emph{piecewise} unless it collapses on the resonant manifold, so I check that
directly at five points before committing to any ansatz.

A pure symmetric polynomial fails on held-out five-point data. That is a clean
negative result: it points to either denominator structure from the internal
propagators or a piecewise dependence coming from the absolute-value kernels. So
I shift from fitting the final amplitude to simplifying the building blocks. The
cubic kernel is already sign-selective, which makes me think a closed form for
the higher kernels could collapse the Berends--Giele result into a manageable
subset-sum expression.

A strong branch then emerges from the data: in the chambers where the
second negative-momentum leg is the smallest scale, the amplitude obeys
\[
A_n \;=\; i\,2^{\,n-1}\,\omega_1\,\omega_2^{\,2n-5}
\]
for $n=5,6,7$. This is encouraging, but it is only one chamber, and I say so
explicitly to myself: I am now \emph{mapping the other chambers to see the
invariant way to state the all-kinematics formula}. Inside each absolute-value
chamber the five-point expression factors cleanly, but there are several
chambers, so I ask Wolfram for a compact absolute-value expression after imposing
the two-minus conservation laws --- which is the right way to get a
chamber-independent statement if it simplifies.

It does not simplify quickly. The symbolic run drags, and then the environment
turns hostile: the shell cannot fork because the machine reports out of memory,
and for a while it refuses to create any new process at all, even a
\texttt{sleep}. So I stop leaning on heavy symbolic simplification and fall back
to small arithmetic checks and to inferring the pattern from \emph{normalized}
numbers. The right normalization is suggested by the leading chamber: I divide
the exact data by $i\,2^{\,n-1}\,h\,s$, where $s$ and $h$ are the two $\sigma=-1$
frequencies with $s$ the smaller in magnitude. The normalized values then depend
only on $U=s^2$ and on the positive-sector squares lying below $U$.

My first compact candidate is to multiply the two negative-sector frequencies by
a \emph{clamped-power} polynomial --- a smooth spline --- in the squared
positive-sector frequencies. The cleanest way to see where it comes from, and
where it breaks, is to track the normalized value as the soft scale $U=s^2$ rises
through the sorted positive squares $x_1\le x_2\le\cdots$, at $n=5$ (so
$m=n-3=2$). My first guess is just the leading chamber --- the bare monomial,
\[
U<x_1:\qquad U^{2},
\]
which is $A_5=i\,2^{4}\,\omega_1\,\omega_2^{\,5}$. As $U$ passes the first knot the
exact data give
\[
x_1<U<x_2:\qquad U^2-(U-x_1)^2 \;=\; x_1\,(2U-x_1),
\]
which is exactly what a single clamped square $U^2-\pp{U-x_1}^2$ predicts --- so
the smooth-spline picture still looks right, and I almost commit. It breaks when
$U$ also passes the second knot. There the smooth clamp predicts
$U^2-\pp{U-x_1}^2-\pp{U-x_2}^2$, but the exact \texttt{BGAmplitude} data give
\[
x_2<U:\qquad U^2-(U-x_1)^2-(U-x_2)^2+(U-x_1-x_2)^2 \;=\; 2\,x_1 x_2 .
\]
The two differ by exactly $(U-x_1-x_2)^2$ --- the ``simple rational amount'' I
kept seeing on a physical $n=6$ point. That missing piece is an overlap:
subtracting the two clamps double-counts the region past \emph{both} knots, and
the $+(U-x_1-x_2)^2$ is the inclusion--exclusion term that restores it. So the
correct object is not a smooth clamped spline at all; it is an \emph{alternating
subset sum}, and I stop trying to bend a continuous function to fit.

Once I read it as inclusion--exclusion the all-$m$ pattern writes itself. At
$n=6$ ($m=3$) the corresponding chamber is
$U^3-(U-x_1)^3-(U-x_2)^3+(U-x_1-x_2)^3$, and the fully-passed chamber collapses to
$6\,x_1x_2x_3=3!\,x_1x_2x_3$. Letting $r$ count how many positive squares lie
below $U$, every chamber is one finite difference over those knots,
\[
G_m\big(U;\{x_j\}\big) \;=\; \sum_{S\subseteq\{1,\dots,r\}} (-1)^{|S|}\,
\Big(U-\sum_{j\in S} x_j\Big)^{m},
\qquad r=\min\big(m,\;\#\{j:x_j<U\}\big),
\]
which is $U^m$ at the $r=0$ end (none passed) and $m!\,x_1\cdots x_m$ at the
$r=m$ end (all $m$ passed), reproducing the two limits the chamber walk forced.
This is the discrete inclusion--exclusion the absolute values were demanding all
along; the smooth clamped spline was the right silhouette but the wrong object.
The amplitude is therefore
\[
\boxed{\;
A_n \;=\; i\,2^{\,n-1}\,g^{\,3-n}\,h\,s\;
G_{\,n-3}\!\big(s^2;\,\{\omega_j^2\}_{\sigma_j=+1}\big)\;}
\]
with the $g$-power fixed by $|k|=\omega^2/g$ and the overall degree $2n-4$
matching the homogeneity I started from. The earlier leading-chamber monomial is
just the $r=0$ end, $G_m=U^m$, where $A_n=i\,2^{\,n-1}\omega_1\omega_2^{\,2n-5}$.

Two things remain. For $n=4$ the manifold is entirely degenerate: a
zero-energy/zero-momentum two-point subcurrent appears and the supplied recursion
returns \texttt{Indeterminate}. I treat it as a limiting value approached from
nearby on-shell external data; the finite continuation of the formula has $m=1$,
so $G_1=U$ and, for example, $\omega=\{-3,2,3,-2\}$ gives $A_4=-192\,i/g$, which I
document explicitly so the reader knows why $n=4$ is a limit rather than a direct
evaluation. And I verify: against the supplied \texttt{BGAmplitude} in exact
rational arithmetic, every nonzero check returns
$\mathrm{Simplify}[\texttt{BGAmplitude}-\text{formula}]=0$, i.e.\ the difference is
identically zero before any floating-point step. The tested points span $r=0$,
the intermediate $r=1,2$, the saturated $r=m$ chamber, mixed signs among the free
frequencies, and a small-frequency regime, for $n=5,6,7$.

I will be honest about the end of the run. After the exact checks the machine
again started refusing new shell processes with an out-of-memory error, so I
could not do a final shell listing or re-run the saved verification script from
the command line; I wrote the result file and the executable harness through the
patch tool instead, and tightened the harness so it resolves \texttt{OnShellBG.m}
relative to its own location rather than the caller's working directory. The
formula and the exact $\mathrm{diff}=0$ checks already collected stand; what I did
not get is one last clean shell rerun.

\subsection{No hint: Fugu ultra}
\label{app:log:case3:fugu}

I can only use the prompt and \texttt{OnShellBG.m}, and I have to be careful not
to read any other pre-existing file. I need a closed-form $A_n$ for
$\sigma=(-1,-1,+1,\dots,+1)$, and I should understand how the kernels work for a
faster implementation. First question: is \texttt{wolframscript} even available?
Let's run \texttt{wolframscript}. It is. I think about how to use the code:
maybe I copy some definitions, maybe I append to \texttt{OnShellBG}, maybe I
write a separate loader that does \texttt{Get}. I write a wrapper that builds the
two-minus sign vector, calls \texttt{MakeKinematics}, and feeds
\texttt{BGAmplitude}, then I just \texttt{Get} the file --- and that was a
mistake: \texttt{Get} runs the file's own demonstration tests, which are for a
\emph{different} sector ($\sigma=(-1,+1,\dots,+1)$, where the amplitude is
identically zero), and it prints a wall of ``$A_5=0$, $A_6=0$, \dots'' while the
$n=8$ case grinds. I let it run for a bit, decide it is wasting time, and
interrupt it. Better idea: import the file as text, split it at the
\texttt{VI.\ TESTS} marker, and \texttt{ToExpression} only the part before that,
so I load the definitions without the demo. (A \texttt{Do[\dots,\{\{n,fw\},\dots\}]}
trips a ``Tag List is Protected'' error too, so I switch the loop to
\texttt{Scan}; small things.)

Now the oracle behaves. The first real surprise is $n=4$: the recursion returns
\texttt{Indeterminate} --- ``$1/0$'', ``$0\cdot\text{ComplexInfinity}$''. The
on-shell point forces an internal two-leg current to zero momentum, so a
propagator blows up and a real cancellation hides behind a removable $0/0$. I set
$n=4$ aside, to come back as a limit, and look at $n=5,6,7$, which give clean,
purely imaginary rationals like $-2304\,i$. So I write $A_n=i\,P_n$ and chase
$P_n$.

I keep thinking about kernels, propagators, and the overall dimensions. Scaling
$\omega\to\lambda\omega$ multiplies $A_n$ by $\lambda^{2n-4}$; changing $g$ shows
$A_n\propto g^{-(n-3)}$; and a family where the second negative leg goes soft,
$\omega_2=a\to0$, shows the amplitude going like $a^5$ when $\omega_2$ is the
smallest scale around. Putting the degree, the $g$-power, and that soft behaviour
together, the simplest thing that fits is the monomial
\[
A_n \;\stackrel{?}{=}\; i\,\frac{2^{\,n-1}}{g^{\,n-3}}\;\omega_1\,\omega_2^{\,2n-5}.
\]
It is suspiciously clean, and it does reproduce the soft data. But it breaks the
instant $\omega_2$ is not the smallest same-sign scale. I notice something
intriguing while poking at this: flipping $\omega_1,\omega_2$ both negative or
both positive gives the same $B\equiv A/(-i)$, so whatever $P_n$ is, it looks
symmetric in the two negative legs. That tempts me to fit a symmetric polynomial
$f(\omega_1,\omega_2)$ of degree six --- $a(u^6{+}v^6)+b(u^5v{+}uv^5)+
c(u^4v^2{+}u^2v^4)+d\,u^3v^3$. I solve for the coefficients on four points; the
solver returns monstrous fractions, and they miss every other point by huge
amounts. So $A_n$ is simply not a polynomial in the two negative frequencies. I
toy with a divided-difference / Lagrange form,
$\sum_i\sigma_i\omega_i^{2n-4}\prod_{j\ne i}\omega_j/(\omega_i-\omega_j)$, and
with B-splines on the sorted (negative) phase velocities, ``normal-form
coefficients based on the max of the frequencies'' --- these feel close to the
right shape but I have no derivation for them yet.

The derivation is hiding in the absolute values. The only non-analytic thing in
the whole construction is $|k_S|=\big|\sum_{i\in S}\sigma_i\omega_i^2\big|/g$
inside each propagator. An absolute value has a corner where its argument changes
sign, so $A_n$ cannot be one polynomial --- it has to be a \emph{spline}, a
piecewise polynomial with break surfaces $\sum_{i\in S}\sigma_i\omega_i^2=0$. That
is exactly why my monomial held in one regime and failed in another: I had been
sitting in one cell and then crossed a wall.

So I stop guessing and resolve the absolute values by hand, chamber by chamber. I
take the symbolic $n=5$ amplitude, collect every \texttt{Abs} argument, replace
each one by $\pm$ itself according to its sign at a sample point in the cell, and
simplify. Inside each cell the expression is an honest rational function, and
after dividing out the common prefactor $i\,2^{4}\,\omega_1\,\omega_2$ the pieces
are, with $x=\omega_2$ and the positive legs $y,z$,
\[
\fitdisplay{
x<y,z:\;\; x^{4},\qquad
y<x<z:\;\; x^4-(x^2-y^2)^2,\qquad
y,z<x:\;\; x^4-(x^2-y^2)^2-(x^2-z^2)^2+(x^2-y^2-z^2)^2 .
}
\]
These are not separate answers; they are the running inclusion--exclusion
corrections of one object. The function ``looks like $x^4-\sum(x^2-y_j^2)_+^2$''.
Writing $t=\omega_2^2$, $a_j=\omega_j^2$, $m=n-3$, and using the truncated power
$\pp{u}=\max(u,0)$, every cell is one value of
\[
T_m\big(t;\{a_j\}_{j\in R}\big)\;=\;\sum_{S\subseteq R}(-1)^{|S|}\,
\pp{\,t-\sum_{j\in S}a_j\,}^{\,m},
\]
the truncated-power / B-spline form I had been circling. The walls
$t=\sum_{j\in S}a_j$ are the $|k_S|$ sign changes, and below all of them only the
$S=\varnothing$ term lives, so $T_m=t^m$ --- which is precisely the principal-chamber
monomial I started from. The whole thing is one truncated power times a universal
prefactor,
\[
\boxed{\;
A_n \;=\; i\,\frac{2^{\,n-1}}{g^{\,n-3}}\;\omega_1\,\omega_2\,
T_{n-3}\!\Big(\omega_2^2;\,\{\omega_j^2\}_{j=3}^{\,n-1}\Big)\;}
\]
with $R=\{3,\dots,n-1\}$ the $n-3$ positive plus-legs.

Two loose ends. For $n=4$ the spline has $m=1$, so the formula predicts
$A_4=i\,(8/g)\,\omega_1\omega_2\min(\omega_2^2,\omega_3^2)$, and I recover the
removable $0/0$ honestly as the $\varepsilon\to0^+$ limit of the
energy-conserving deformation
$\omega=\{-\omega_3-\varepsilon,\omega_2,\omega_3,-\omega_2+\varepsilon\}$; the
limit is finite and gives $-192\,i$ for $\{2,3\}$, matching. The other loose end
is which leg is ``$t$'', and here I genuinely stumble: my first channel guess
predicts $-3776$ where the oracle says $-701$ for one point, and it is fine for a
small $\omega_2$ but goes wrong once $\omega_2$ is not small. I think the
discrepancy is about signs, and I reframe it as a channel-selection question:
counting left- and right-movers, my form works for one negative leg against
$n-3$ same-sign plus legs, and the right rule is to let $p$ be the $\sigma=-$ leg
whose sign is shared by exactly $n-3$ of the plus legs (the set $R$), with $q$ the
other $\sigma=-$ leg, so
$A_n=i\,2^{\,n-1}q\,p\,T_{n-3}(p^2;\{\omega_j^2\}_{j\in R})/g^{\,n-3}$.

My first script for this channel-aware form fails on every line --- but the
failure is a \texttt{Function::slotn} slot mistake in my selector code, not a
disagreement with the oracle. I fix the selector and rerun, and now the closed
form reproduces \texttt{BGAmplitude} \emph{exactly}: the relative error is
identically zero in exact rational arithmetic for $n=4,5,6,7$, at many points per
$n$, including non-generic regimes where one frequency is a hundred times larger
or smaller than the others, and at $g\ne1$.

I want to be honest about the scope rather than overclaim. For arbitrary positive
free frequencies handed to \texttt{MakeKinematics} --- the setup the prompt
actually specifies --- energy conservation drives the two solved legs negative
and the kinematics always land in this one physical channel; I check five hundred
random points for each $n=4,\dots,8$ and the boxed formula is exact on all of
them. If instead one hand-flips the signs of the free frequencies into other
cells of the spline, the same truncated-power structure persists but with a
different leg as $p$ and a different same-sign set as $R$; I state that selection
rule but I did not fold every sign cell into a single universal expression. So,
honestly: within the prompt's own kinematics this is the closed-form answer and
it is verified to exact equality; the fully sign-general statement I leave as the
channel-selection rule plus the spline structure, not as one compressed formula.

\bibliographystyle{plainnat}
\bibliography{references}

\end{document}